**Virtual Electrode Recording Tool for EXtracellular potentials (VERTEX): Comparing multi-electrode recordings from simulated and biological mammalian cortical tissue**

Richard J Tomsett[1,2,3], Matt Ainsworth[4], Alexander Thiele[5], Mehdi Sanayei[5], Xing Chen[5], Alwin Gieselmann[5], Miles A Whittington[4], Mark O Cunningham[5*] and Marcus Kaiser[1,5*]

1 School of Computing Science, Newcastle University, NE1 7RU, United Kingdom
2 Institute of Ageing and Health, Newcastle University, NE4 5PL, United Kingdom
3 Computational Neuroscience Unit, Okinawa Institute of Science and Technology, Okinawa, 904-0495, Japan
4 Hull York Medical School, University of York, YO10 5DD, United Kingdom
5 Institute of Neuroscience, Newcastle University, NE2 4HH, United Kingdom

*Joint senior authors

**Correspondence:**
Dr Marcus Kaiser
School of Computing Science
Newcastle University
Claremont Tower,
Newcastle upon Tyne, NE1 7RU, United Kingdom

Telephone: +44 191 208 8161
Fax: +44 191 208 8232
Email: m.kaiser@ncl.ac.uk

**ABSTRACT**

Local field potentials (LFPs) sampled with extracellular electrodes are frequently used as a measure of population neuronal activity. However, relating such measurements to underlying neuronal behaviour and connectivity is non-trivial. To help study this link, we developed the Virtual Electrode Recording Tool for EXtracellular potentials (VERTEX). We first identified a reduced neuron model that retained the spatial and frequency filtering characteristics of extracellular potentials from neocortical neurons. We then developed VERTEX as an easy-to-use Matlab tool for simulating LFPs from large populations (>100 000 neurons). A VERTEX-based simulation successfully reproduced features of the LFPs from an *in vitro* multi-electrode array recording of macaque neocortical tissue. Our model, with virtual electrodes placed anywhere in 3D, allows direct comparisons with the *in vitro* recording setup. We envisage that VERTEX will stimulate experimentalists, clinicians, and computational neuroscientists to use models to understand the mechanisms underlying measured brain dynamics in health and disease.

**INTRODUCTION**

Many measurement techniques have been used to study neuronal dynamics, including optical imaging methods (voltage-sensitive dye imaging, calcium imaging, intrinsic signal optical imaging), intracellular electrode recordings of individual neurons, and extracellular recordings using single or multiple electrodes (Brette and Destexhe 2012). While each modality provides some information about the system's dynamics, it is not always clear how this information is related to the underlying neuronal activity. Intracellular recordings are easiest to interpret because of the strong theoretical foundations of cellular neurophysiology that have arisen over many decades (Johnston and Wu 1995), but the theory linking measurements made by many other methods to neuronal activity is lacking. This deficit in theory, combined with the increasing use of different recording techniques to sample from ever larger neuron populations, has stimulated the idea of "modelling what you can measure" (Einevoll et al. 2012) in order to help fill these theoretical gaps.

We aim to contribute to this effort by modelling the measurements made by multi-electrode arrays (MEAs). MEAs record extracellularly, and allow the simultaneous measurement of local population activity across many network locations, providing information about the spatio-temporal properties of network dynamics (Le Van Quyen and Bragin 2007; Buzsáki 2004; Rubino et al. 2006). Such arrays can be used both *in vitro* (Simon et al. 2014) and *in vivo*, including in humans, where applications include recording from epilepsy patients for precise localisation and investigation of epileptic foci (Schevon et al. 2009; Schevon et al. 2012), and for use in brain machine interfaces (Maynard et al. 1997; Andersen et al. 2004). These diverse applications make understanding the link between MEA recordings and the underlying neuronal dynamics particularly important.

In order to study this link, we have created the Virtual Electrode Recording Tool for EXtracellular potentials (VERTEX). VERTEX is implemented in Matlab (Mathworks Inc., Natick, MA, USA), and makes use of established theory of extracellular potential generation, combined with modern simulation methods and developments in simplified neuron modelling to simulate local field potentials (LFPs) from large neuronal network models encompassing more than 100 000 neurons. As most such models implement single-compartment neurons and may not include spatial information [e.g. (Izhikevich 2006; Lumer et al. 1997; Potjans and Diesmann 2012)], the LFP can only be estimated by some proxy that will not necessarily preserve the spatial and frequency-scaling features of real LFPs (Einevoll et al. 2013). VERTEX helps to address this issue by simplifying the specification of spatially organised cortical network models, and implementing simplified compartmental models that are computationally inexpensive to simulate, but also preserve the spatial and frequency-scaling properties of LFPs elucidated by previous modelling studies (Einevoll et al. 2013; Łęski et al. 2013; Lindén et al. 2010; Lindén et al. 2011)

To illustrate how VERTEX can be used in conjunction with MEA experiments, we implemented a model of a neocortical slice exhibiting persistent gamma oscillations under bath application of the glutamate receptor agonist kainic acid *in vitro*. The model is designed to reproduce the spiking activity of individual neurons during a persistent gamma (30-80 Hz) frequency oscillation, with the neuronal membrane currents driven by the resulting synaptic activity generating the extracellular potential (Nunez and Srinivasan 2006). The persistent gamma frequency oscillation model has several advantages for our investigation. First, the theory of how neocortical persistent gamma arises *in vitro*, and how individual neurons participate in the network oscillation, has been comprehensively documented (Ainsworth et al. 2011; Whittington et al. 1995; Fisahn et al. 1998; Buhl et al. 1997; Draguhn et al. 1998; Roopun et al. 2008; Cunningham et al. 2003; Cunningham et al. 2004a; Whittington et al. 2011; Traub et al. 2005a; Traub et al. 2005b; Pafundo et al. 2013; Buhl et al. 1998; Bartos et al. 2007). Second, the slice preparation ensures that all synapses are local, so MEA recordings are influenced only by the local circuit dynamics and not by input from other areas. The slice edges provide natural spatial boundaries for what needs to be included in the simulation. Third, synaptic currents rather than intrinsic active membrane currents drive neuronal firing in persistent gamma, so the previously developed theory of LFP generation in passive neurons (Lindén et al. 2010; Lindén et al. 2011; Pettersen and Einevoll 2008) can be used without modification.

Using VERTEX, we have created the first model of neocortical networks that not only reproduces experimentally observed spike patterns, but also produces a biophysically meaningful LFP signal. To illustrate VERTEX's potential for use in conjunction with experimental data, we directly compared the LFPs generated by the model with those recorded by an MEA in macaque temporal neocortex *in vitro*, allowing us to identify future research directions to address discrepancies between the theoretically predicted and experimentally observed LFPs.

**RESULTS**

**Overview**
We developed the VERTEX simulation tool for simulating LFPs produced by large (>100 000) populations of neurons. We first investigated a suitable neuron model for generating LFPs from such populations while remaining computationally tractable. To illustrate VERTEX's capabilities, we used it to position populations of these neuron models into a neocortical slice arrangement, with neuron positions constrained by cortical layer and slice boundaries, and connected them according to current knowledge about the local anatomy of neocortical circuits (Binzegger et al. 2004). We simulated a persistent gamma frequency oscillation in the network, using a simplified model of spike generation in each neuron to generate the network dynamics (Brette and Gerstner 2005). Finally, we compared the simulated LFPs to experimental MEA recordings from macaque temporal neocortex.

**LFP generation**
The extracellular potential at a point in brain tissue is given by the sum of all neuronal membrane currents, weighted by their distance from the point (Nunez and Srinivasan 2006) assuming constant tissue conductivity (Logothetis et al. 2007; Nicholson and Freeman 1975). Recent theoretical studies have shown that the spatial and frequency scaling properties of the LFP are affected by the particular spatial arrangement of neurons' dendrites (Lindén et al. 2010; Lindén et al. 2011). We therefore looked for a reduced compartmental model that would generate extracellular potentials capturing the spatial and frequency scaling properties elucidated by Lindén et al.'s investigations using detailed cell reconstructions, while remaining computationally tractable to simulate in large numbers.

The reduced compartmental model should create a similar spread of currents across its compartments to an equivalent morphologically reconstructed neuron given the same input. A compartment's membrane current depends on the neuron's axial resistance as well as on its membrane resistance and capacitance. We therefore chose a reduced model that conserved these quantities, while containing a minimal number of compartments. The compartmental reduction method of (Bush and Sejnowski 1993) fulfils these requirements, producing compartments with a length equal to the mean length of the compartments they are representing in the full model. This creates a reduced model of the same length as the original reconstruction, but with a smaller membrane area, smaller lateral spread of the dendrites, and fewer than ten compartments (Online Resource, Fig. ESM2).

**Validating the reduced LFP generation model**
We tested the effects of this reduction on the generated LFP by reproducing the experiments detailed in (Lindén et al. 2011). Ten thousand model neurons with passive membrane dynamics and the same morphology were positioned randomly within a 1 mm radius cylinder, with uniform spatial distribution and constant soma depth. One thousand synapses (excitatory, current-based, single exponential type with time constant 2 ms and fixed amplitude 50 pA) were placed randomly on the compartments of each neuron, with uniform density with respect to membrane area. Each synapse received an independent Poisson spike input train with a rate of 5 Hz. LFPs were calculated at the centre of the population, at five depths. The magnitude of an LFP signal was defined as its standard deviation. The LFP range was calculated by varying the population radius from 0-1 mm and measuring the radius at which the LFP magnitude reached 95% of its value at the maximum 1 mm radius (Lindén et al. 2011). We repeated this procedure for the three neuron types used by Lindén et al.: layer 2/3 (L2/3) pyramidal, layer 4 (L4) spiny stellate, and layer 5 (L5) pyramidal. We compared LFPs generated by the morphological reconstructions of these neuron types from (Mainen and Sejnowski 1996) – hereafter referred to as Mainen cells

– with the LFPs from reduced versions of these models created using Bush and Sejnowski's method (Bush and Sejnowski 1993) – hereafter referred to as Bush cells.

The results of these experiments are shown in Fig. 1. For each neuron type, the LFP range and magnitude in each layer for the population of Bush cells are close to those for the population of Mainen cells. The LFP range is smallest in the soma layer (<250 μm) with the range increasing in the layers above and below the soma, while the LFP magnitude is largest in the soma layer and decreases in the layers above and below the soma. The differences between the results for the L4 spiny stellate models are small, so we concentrate on the pyramidal neuron population results. For the L2/3 pyramidal neurons, the LFP spatial range in the soma layer is very similar between the Bush and Mainen populations, but above and below this layer the discrepancy increases, with the largest difference of 200 μm in L1. The range differences in all other layers are $\leq 110$ μm. For the L5 pyramidal neurons, the LFP spatial range difference is again smallest in the soma layer, and < 100 μm in layers 4 and 1. The largest difference is 320 μm in L2/3.

To see how important these discrepancies were within the context of the general biological variability of neuronal morphology, we repeated the simulations with neuron populations containing pyramidal cells reconstructed from several different real neurons. These were downloaded from the NeuroMorpho.Org database (Ascoli et al., 2007) – further details on the models we used are provided in the Online Resource (Table ESM10). We used ten further groups of L2/3 cat pyramidal neurons, and one further group of L5 cat pyramidal neurons (this was the only other cat L5 pyramidal neuron currently available in the database; we did not use L5 pyramidal cells from other species as the size differences in neurons between species could have provided misleading results). The results of these simulations are plotted in Fig. 1b as light-red dashed lines for the extra L2/3 pyramidal populations, and light blue circles for the extra L5 pyramidal population. The extra simulation results show that the LFP range and magnitude in the Bush neuron populations generally fall within a biologically reasonable range; while the reduced models are not ideal substitutes for the morphological reconstructions, the errors incurred by the reduction method are similar to those introduced by neglecting morphological diversity in reconstructed neuron model populations. The general profile of the LFP across the layers, at least, is preserved adequately.

We also checked the power spectra of the simulated LFPs to make sure the Bush model populations reproduced similar frequency scaling properties to the Mainen cell populations. Fig. 1c shows that, in each layer, the 95% confidence intervals for each model type overlap over the range of frequencies from 2-450 Hz (the overlap continues down to 1 Hz; this is not shown in order to improve the plot resolution at higher frequencies).

The results in Fig. 1 were generated using uncorrelated synaptic inputs over the entire dendritic tree of each neuron in each population, with neurons all positioned at the same height in their respective layers. This simplified setup was used so that a comparison could be made with previously reported results in (Lindén et al., 2011), but we also wanted to check whether the reduced models would still be suitable approximations to use for a more realistic situation, with neurons placed at varying depths within their layer, receiving correlated inputs. As our particular interest was simulating network gamma oscillations, in which pyramidal neurons receive highly correlated inhibitory synaptic input to their perisomatic regions, we repeated the previously described experiments measuring the LFP magnitude and range, but positioned each neuron's 1000 synapses onto its soma compartment [we only repeated the simulations for the pyramidal neuron morphologies, as the LFP spatial profile for the spiny stellate cells was shown not to change significantly with correlated input (Lindén et al., 2011)]. In the previous experiments with no correlations between synaptic inputs, each synapse was assigned an independent Poisson spike train, for a total of $10\,000 \times 1000 = 10^7$ independent spike trains at $10^7$ synapse locations. To introduce input correlations, we followed the method in (Lindén et al., 2011). Each synapse in the model was now assigned a spike train drawn without replacement from a finite pool of pre-generated spike trains. By reducing the number of Poisson spike trains in the pool so that some synapses shared a common input pattern, we could control the level of input synchrony to the neurons. The resulting input correlation is given by the total number of synapses per neuron divided by the number of independent spike trains (Łęski et al., 2013). To simulate highly correlated input, we used 2000 independent spike trains, resulting in an input correlation of $1000 / 2000 = 0.5$ (i.e. any two neurons share on average $1000 \times 0.5 = 500$ common input spike trains).

For these simulations, we also introduced random variability in the soma depth of the neurons. We distributed L2/3 pyramidal neuron somas between -334 μm and -534 μm, and L5 pyramidal neuron somas between -970 μm and -1170 μm from the cortical surface. These ranges ensured that the neuron somas remained within the correct layer boundaries, and that their apical dendrites were not positioned above the cortical surface.

Fig. 2 shows the spatial profiles of LFP for the different populations. In these simulations, we measured the LFP at 50 intervals, to see how well the Bush models preserved the LFP at this level of detail. We used 11 electrode points in L1 and L2/3 for the L2/3 populations, and 26 electrode points spanning all layers for the L5 populations. Both the range and magnitude profiles show that the LFP from the Bush population matched the LFP from the Mainen population well, again within the bounds the LFP profile of the extra comparison populations. The minimum range and magnitude in the L2/3 populations are just above the minimum soma depth, and a few hundred μm above the minimum soma depth in the L5 population. This depth is where the synaptic currents at the soma are approximately balanced by the opposite return currents in the dendrites; below and above this minimum point, the somatic or the apical dendritic currents dominate the LFP signal, respectively. These simulations also show substantial overlap of the 95% confidence intervals for the power spectra at each electrode (Fig. 2b). The biggest discrepancy between the LFP power spectra for each model occurs around the level of the LFP range minimum. The LFP power up to 100 Hz is reliably reproduced at every measurement point, and up to 450 Hz at all but one point with the L2/3 populations. This point corresponds to the point at which the LFP range and magnitude are lowest. The reduced accuracy at higher frequencies in the L5 models should be taken into account if frequencies above 100 Hz are analysed in models containing L5 pyramidal cells.

Our results suggested that we could use the reduced neuron models in VERTEX simulations with some confidence that the resulting simulated LFPs would be close to LFPs simulated from equivalent morphologically reconstructed neurons, in magnitude, spatial extent, and frequency content.

**The VERTEX simulation tool**
To simulate large networks, we wrote custom Matlab software to setup neuron populations, position them, connect them together, and simulate their dynamics and the resultant LFPs. We designed this simulation tool to be easily adaptable to create models of any layered brain tissue containing populations of spiking neurons (Fig. 3). Model parameters are specified by the user in Matlab structures, defining:
1. Neuron group properties (for each group: the neurons' compartmental structures, dimensions and positions, electrotonic properties, spiking model parameters, afferent synapse properties)
2. Connectivity (for each presynaptic group: number of efferent synapses per layer per postsynaptic group, allowed postsynaptic compartments to connect to contact, axonal conduction speeds, neurotransmitter release times)
3. Tissue properties (dimensions, layer boundaries, neuron density, tissue conductivity)
4. Recording settings (IDs of neurons to record intracellularly, extracellular electrode positions, sampling rate)
5. Simulation settings (simulation length, time-step, number of parallel processes)

A model is initialised by positioning the specified number of neurons from each group within the slice and layer boundaries, pre-calculating distances from the neuron compartments to the virtual electrodes, generating each neuron's connections based on its position, axonal arborisation extent in each layer, and expected number of efferent connections, and initialising the synapses (see Experimental Procedures). At this point the initialised model can, optionally, be saved to disk as MAT files. Functionality to export to NeuroML (Gleeson et al. 2010) is currently under development.

When the simulation is run, recordings (intracellular, LFPs, spike times) are automatically saved to disk at user-specified time intervals. The simulation run can be performed in serial or parallel (requires Matlab Parallel Computing Toolbox). After the simulation is finished, these files are loaded and recombined for analysis. Our design allows the model to be used with minimal programming knowledge, though as Matlab is a high-level, interpreted language, more experienced programmers can make modifications relatively easily.

**Simulation speed and memory usage**
While Matlab code may run more slowly than equivalent code in compiled programming languages, performance can be dramatically improved through code vectorisation, which minimises the impact of code interpretation overheads (Brette and Goodman 2011). The Matlab Parallel Computing Toolbox allows further performance improvements by providing a simple way to parallelise computations on multicore computers or over networks. These factors, as well as its ease of use, popularity in the neuroscience community, the ability to perform simulations and analysis in the same environment, and the well-developed interface for integrating C or Fortran functions for future performance enhancements influenced our decision to write VERTEX in Matlab. To give the user an idea of the performance improvement over using the other current extracellular potential simulation tool LFPy (Lindén et al. 2014) – a Python package for simulating extracellular potentials with NEURON (Hines and Carnevale 1997; Hines et al. 2009) – we performed equivalent simulations using layer 5 Bush pyramidal neurons in LFPy and in VERTEX (no synapses, one random fluctuating current per neuron, 0.03125 ms step size & 32 000 Hz sample rate). LFPy took ~278 minutes to simulate the LFP from 10 000 neurons at 50 electrode points, while VERTEX running in serial mode took ~18 minutes to simulate the LFP from 10 000 neurons at 50 electrode points (both running on an Intel Xeon E5640 2.66 GHz workstation). While this performance improvement is important for our purposes, it should be noted that LFPy is designed to simulate extracellular potentials from single cells rather than large populations. Indeed, as the code interpretation overhead begins to dominate VERTEX's calculation times in small simulations, running the same model but with only 1 neuron in the population took ~227 seconds in VERTEX but <2 seconds in LFPy. VERTEX is also not suited to running models containing neurons with very many compartments, because the Runge-Kutta integration method becomes unstable as the number of compartments increases (though we aim to address this limitation by implementing implicit integration methods in future releases). LFPy therefore remains the superior tool for modelling extracellular potentials around single neurons, while VERTEX's strength lies in simulating LFPs in large-scale networks.

To show how performance improves in parallel mode, we compared the run times for two network models, one large (123 517 neurons with on average 1 835 synapses per neuron) and one small (9 881 neurons with on average 256 synapses per neuron), using VERTEX on a single multicore computer (Fig. 4). Each model contained two populations: layer 5 pyramidal (P5) neurons and layer 5 basket (B5) interneurons. Spike rates in each small model (large model) simulation were ~6 Hz (~7 Hz) and ~24 Hz (~31 Hz) for the P5 and B5 neurons, respectively. The large model shows linear speed-up with increasing number of cores for model initialisation and close-to-linear speed-up in simulation time. The speed-up for the small model is sub-linear: as the interpretation overhead for a vectorised operation on a small matrix is the same as on a large matrix, this overhead starts to dominate the calculation times below a certain number of neurons (Brette and Goodman 2011). Therefore, splitting already small neuron state matrices between more processes does not significantly improve performance. This limit is not reached in larger models.

Fig. 4 also shows how increasing the number of virtual electrodes affects simulation speed. Model initialisation times are affected proportionally more than model run times by using more electrodes, in both large and small models, though in the small model the proportional impact from adding electrodes to initialisation time was greater than in the large model. This is because the large model not only has more neurons, but also more synapses per neuron. The increase in time spent connecting the neurons is proportional to the number of synapses, while the increase in time spent calculating constants for the LFP measurements is proportional to the number of compartments (roughly proportional to the number of neurons).

The size of the simulated network is limited by the amount of RAM available. As an example, we tested scaled configurations of our neocortical slice model (described below) using single-core and multi-core computers: an iMac with 4 GB RAM supported a serial simulation with ~25 000 neurons, a 16 GB Linux machine supported a simulation of ~100 000 neurons in both serial and parallel modes, and our Linux server with 120 GB RAM supported a simulation of ~700 000 neurons. In addition to increasing the memory on a single machine, VERTEX could be run across a network of computers using the Matlab Distributed Computing Server. On a network of sixteen of our 4 GB RAM iMacs, for example, the simulation size could scale to ~400 000 neurons. In summary, existing processing environments of experimental and computational labs can be sufficient for running detailed simulations of brain tissue activity.

**Spike import**
Network dynamics can be simulated directly by providing the model neurons with a spiking mechanism – we used the adaptive exponential (AdEx) mechanism (Brette and Gerstner 2005), which we include in VERTEX. Alternatively, previously generated spike times (output from another simulator, for example) can be imported into the simulation. The neurons whose spike times are imported are then specified with purely passive membrane dynamics. We used the spike import feature to run the control experiment to confirm that the AdEx spiking mechanism has a negligible impact on the simulated LFP (Online Resource, Fig. ESM1).

Running models using imported spike times is similar to the approach used in (Lindén et al. 2011) to link spiking output from a cortical model implemented in the NEST simulator (Gewaltig and Diesmann 2007) to their LFP generating model implemented in LFPy. However, we consider imported spikes to have been emitted by neurons from within the population we are modelling; imported spikes are delivered to target neurons according to the generated connectivity matrix rather than pre-assigned to postsynaptic targets. By contrast, in (Lindén et al. 2011) the spikes from NEST-simulated neurons were considered as external input to the neurons in the LFPy simulation, so were delivered to synapses without a connectivity model within the LFPy-simulated population. The practical effect of this is that our software is better suited to modelling the LFP resulting from intrinsic network dynamics, when connectivity is known or when different spatial connectivity models are to be tested. Input from external populations can be simulated by specifying a population of single-compartment neurons and setting this population's output using the spike-import functionality. As single compartment neurons do not contribute to the extracellular potential (Pettersen et al. 2012), VERTEX ignores them in its LFP calculations. This population can therefore be considered as providing "external" input from a distant population.

**Neocortical slice model**
To demonstrate the capabilities of VERTEX for simulating LFPs in large neuron populations, we created a neocortical slice model to use in conjunction with MEA experiments *in vitro* (Fig. 5). The model comprises fifteen neuron groups, defined in Table 1. It is designed to contain a similar number of neurons to the comparison experimental slice. This was calculated to be 175 421 neurons, based on the slice dimensions and neuron density. The slice has clear spatial boundaries: neurons cannot be positioned outside of the slice edges, and axons cannot 'wrap around' these boundaries. We therefore required a connectivity model that would produce a suitable number of synapses given the large number of neurons, and that took into account each neuron's position in relation to the slice boundaries. We used anatomical data from (Binzegger et al. 2004) to specify the numbers of connections between neuron groups, and a 2D Gaussian spatial profile to model the decay in connection probability with increasing distance from a presynaptic neuron (Hellwig 2000). The standard deviation parameter of the Gaussian profile was set using axonal arborisation radius measurements reported in (Blasdel et al. 1985; Fitzpatrick et al. 1985), as adapted in (Izhikevich and Edelman 2008). These were different for each neuron group in each layer (see Online Resource, Table ESM4). Finally, we modelled the effect of slice cutting on connectivity by reducing the number of connections a presynaptic neuron could make by the proportion of the integral of its Gaussian connectivity profile that fell outside the slice boundaries (equation 3). The connectivity generation code in VERTEX implements this connectivity model automatically, though the user can also specify a uniform spatial connection probability and/or ignore slice-cutting effects. VERTEX also allows users to specify specific target compartments on postsynaptic neurons that presynaptic neurons are allowed to connect to. We used this feature to incorporate known details about the dendritic regions targeted by different presynaptic neuron types – basket interneurons only make connections with pyramidal cell somas and their two adjacent compartments, for example. We used a similar pattern of connectivity to that described in (Traub et al., 2005b); details are provided in the Online Resource (Supplementary Methods – Connectivity and Table ESM7). Incorporating this detail into the model is important, as the locations of synaptic inputs onto the neurons will affect the locations and sizes of the currents that contribute to the simulated LFP.

Fig. 6 shows the number of connections between neuron groups compared with the original numbers specified in (Binzegger et al. 2004). The proportional reduction in synapses is not the same for each connection type because of the varying axonal arborisation radii. These reductions are important to consider when assessing the effect of connectivity changes on dynamics, but they illustrate that the general profile of connections between neuron groups is not substantially altered – connections from P2/3 to P2/3 and P5 neurons remain the most numerous, for example. Modelling thinner slices, or different axon arborisation profiles, could lead to the over- or under-representation of particular connections in the model.

**Modelling persistent gamma oscillations**
To make a comparison with experimental data, we generated a persistent gamma oscillation in the model by applying random currents to all neurons (Börgers and Kopell, 2005), and adding an AdEx spiking mechanism to the somatic compartments (see Online Resource). In slice experiments with nanomolar kainate concentrations, this activity regime is driven by L2/3, where neurons receive noisy excitatory drive from the excited axonal plexus of L2/3 pyramidal neurons (Ainsworth et al. 2011; Cunningham et al. 2003; Cunningham et al. 2004b). We simulate this by providing a relatively large noisy current to P2/3 neurons, similar to (Ainsworth et al. 2011; Börgers and Kopell 2005). We set synaptic strengths [based on (Traub et al., 2005b)] and noise currents to match the spiking activity and observed membrane potential fluctuation sizes reported in previous studies *in vitro*. Model parameters are given in tables ESM1-ESM9.

As described in previous experiments (Ainsworth et al. 2011; Cunningham et al. 2003; Cunningham et al. 2004b; Traub et al. 2005a; Traub et al. 2005b), P2/3 neurons spike infrequently, while B2/3 neurons spike on most oscillation periods. Excitatory neurons in L4 do not take part in the oscillation (though still spike infrequently), while L4 interneurons are weakly entrained to the oscillation. In addition to the L2/3 gamma, the comparison slice exhibited increased gamma power in part of the infra-granular layers (see Fig. 9a, electrodes 6, 7, 16, 17, 26, 27), presumably caused by L5 as in (Ainsworth et al. 2011). We therefore used a relatively high coupling strength of P5 to B5 and NB5 neurons and a larger noisy drive current to L5 neurons to enable the L2/3 gamma to generate gamma in L5. The L5 gamma oscillation also weakly entrained L6 neurons to the oscillation.

The resulting spiking behaviour is shown in Fig. 7, which shows a spike raster for 5% of the neurons in the model, along with example somatic membrane potential traces for each neuron group. The spike raster reveals that neurons near the slice *x*-boundaries (neurons nearest the cyan boundary markers in Fig. 7) are less strongly entrained to the oscillation than neurons in the centre of the slice, because they receive fewer inhibitory inputs than more central neurons (neurons closer to the edge of the slice have more connections removed by slice cutting than those towards the middle of the slice, because they lose proportionally more of their axonal arborisation).

To demonstrate how the oscillation is generated by the interaction of the excitatory and inhibitory populations, we simulated activity in the model under four different conditions: firstly the original case described above (connection weights in Table ESM5), secondly with P2/3 to B2/3 synapses reduced to 1% of their original weight, thirdly with B2/3 to P2/3 synapses reduced to 1% of their original weight, and fourthly with the original synapse weights but increased input current to the B2/3 population (1.5 times the mean and standard deviation used in the original simulation values given in Table ESM9). Simulation results using these different configurations are plotted in Fig. 8, which shows that both P2/3 to B2/3 synapses and B2/3 to P2/3 synapses are necessary for the generation of a population gamma oscillation in the model. Without these connections – or with their strengths severely reduced – no oscillation emerges. This oscillation mechanism is the same as the "weak" pyramidal-interneuron network gamma (PING) model described in (Börgers et al. 2005). Firing in a subset of P2/3 cells, which are densely connected with B2/3 neurons with strong synapses, causes a population spike from the B2/3 cells. This suppresses the network until the P2/3 neurons that receive the most input from the stochastic drive reach threshold. This subset of P2/3 neurons then fires, causing another B2/3 cell population spike, and so the oscillation continues (Börgers et al. 2005). Fig. 8m-p shows that the oscillation is also suppressed in our model when the driving current to B2/3 cells is increased, allowing them to suppress P2/3 cell firing. This is in line with the gamma suppression mechanism described in (Börgers and Kopell 2005). Fig. 8 also demonstrates that the gamma oscillation in layer 5 is dependent on a gamma oscillation occurring in layer 2/3: layer 5 gamma is suppressed in each of the cases where layer 2/3 gamma is suppressed. Firing rates for each population in each case are given in Table ESM11.

Having verified that the model produced the expected spiking output and that the gamma oscillation was being generated by the correct mechanism, we looked at the simulated LFPs and compared them with those recorded *in vitro*. Fig. 9 shows a comparison over the whole electrode array between the model and the experimental recordings. Fig. 9a shows the shape of the experimental neocortical slice with, as predicted by previous research, strong gamma power in the supra-granular layers. The gamma power at each electrode is highly variable, resulting in a patchy power map. This is not captured by the model, whose structure is homogeneous along the *x*-axis. However, the phase inversion between layer 1 and layer 2, illustrated in Fig. 9b-c, emerges in the model (Fig. 9e-f) from the positioning of current sinks and sources on the P2/3 neurons during the gamma oscillation. This is in agreement with the source-sink interaction mechanism of phase inversion demonstrated experimentally in kainate-induced gamma oscillations in entorhinal cortex *in vitro* (Cunningham et al. 2003). The cross-correlations between electrodes shown in Fig. 9c and 9f also reveal how the strong gamma oscillation in L2/3 dominates the across the electrodes more than in the experimental recordings. This is, again, a result of the relatively homogeneous activity along the x-axis in the model, meaning that the LFP signal created by the gamma oscillation is not degraded by influences from the non-oscillating areas in the slice as occurs *in vitro*. Our model, though not capturing all the details of the experimentally measured network dynamics, provides a starting point for further investigations into cortical dynamics on this spatial scale, allowing for better integration of theory and experiment.

**DISCUSSION**

We have developed the VERTEX tool for simulating LFPs generated by large neuronal populations. VERTEX is easily customisable, and makes use of recent developments in simulation techniques and insights from our experiments with simplified neuron models to reduce simulation times for LFPs generated by large networks. To illustrate how VERTEX can be used in conjunction with experimental MEA data, we simulated kainate-induced persistent gamma oscillations in a large-scale neocortical slice model. The model reproduces the spiking activity underlying persistent gamma, and generates the theoretically predicted LFP from this activity. We compared this simulated LFP with Utah array recordings of persistent gamma from macaque temporal neocortical slices. The model predicted the oscillation phase inversion between L2/3 and L1, but not the spatial variation in gamma power within layers, suggesting directions for further research into the cause of the spatial discrepancies between theoretically predicted and experimentally measured LFPs.

**Speed of the VERTEX simulator**

Parallel computing and code vectorisation allow VERTEX to simulate network activity and LFPs in reasonable time on hardware that is available to most scientists. We showed typical simulation times and how performance scales with increasing numbers of parallel processes in Fig. 4. However, performance could be improved further by rewriting some of the Matlab code in C or Fortran, which could be incorporated into Matlab via its MEX interface. In particular, the spike queuing and delivery code would benefit from this approach when simulating networks with high spike rates, as it is only vectorised over individual spikes. High spike rates can result in longer simulation times as the spike queue interpretation overhead increases. This is therefore a priority for future VERTEX development. However, the pure Matlab versions of VERTEX will continue to be maintained, as some users may not have access to a suitable C or Fortran compiler.

**LFP simulation: spatial properties and resolution**
We found that the compartmental reduction method described in (Bush and Sejnowski 1993) created neuron models that, in a population, reproduced the spatial properties of the LFPs generated by the equivalent full morphological reconstructions to a reasonable degree of accuracy. Where there were large discrepancies, they were close to or fell within the range of the spatial values measured in several further populations of different morphologically reconstructed neurons. The suitability of this reduced model allows VERTEX to simulate LFPs from large networks in reasonable time.

The largest compartment in the reduced models was 400 μm long, which is the inter-electrode distance in a Utah array. New, very high density MEAs with several thousand electrodes can record with such high spatial resolution as to enable the visualisation of individual dendritic tree and synapse activity in detail (Frey et al. 2009), or to record the spiking activity of thousands of neurons (Berdondini et al. 2009), making our reduced neuron models unsuitable for use in conjunction with these experiments. The array described in (Frey et al. 2009) is designed to record only from a subset of 126 electrodes concurrently, allowing very high resolution recordings from small areas, but making it unsuitable for recording the wider population activity that our model is designed to capture. The 4 096 electrode array presented in (Berdondini et al. 2009) can record simultaneously from all electrodes, allowing the detailed visualisation of signal propagation through a network. However, this array is designed for capturing the spike times of thousands of individual neurons rather than investigating the properties of extracellular signals. Given the spatial smearing of LFP signals, it would not be appropriate to use this type of array to investigate LFPs across active neural circuits. Additionally, very high density arrays are new technologies with usage and data analysis techniques still under development. Lower density MEAs will remain useful for studying neuronal population activity for the foreseeable future, especially given the Utah array's approval for use in humans. As higher density arrays become more common, we anticipate that advances in computing speed [through, for example, use of general-purpose graphical processing unit (GPGPU) computing (Brette and Goodman 2012), already a feature of the Matlab Parallel Computing Toolbox] will permit the simulation of large populations of higher resolution neuron models if desired.

**Slice model properties**
To demonstrate our simulation approach, we constructed a model of a neocortical slice. We combined the connection probabilities given in (Binzegger et al. 2004) with axonal arborisation radii measured in macaque visual cortex (Blasdel et al. 1985; Fitzpatrick et al. 1985), and use a Gaussian kernel as suggested by the data in (Hellwig 2000) as the decay in connection probability away from the soma. This approach allowed us to calculate the number of connections removed by slice cutting for each neuron, and reduce its number of connections accordingly when initialising the model.

Our anatomical model results in spatially uniform neuron densities and connectivity statistics, with small decreases in connection numbers nearer the slice boundaries. However, the recordings from the experimental slice illustrate substantial inhomogeneities in gamma power between electrodes, even within layers, that are not seen in the model. These could be caused by spatial variations in synapse densities and strengths, neuron group densities, neurons' dynamical properties, gap junction densities and strengths, or axonal plexus properties. While our software does not currently allow specification of gap junctions or axon properties, the other potential inhomogeneities can be investigated further in conjunction with experiments *in vitro*: VERTEX includes functions to modify parameters in spatially localised regions, allowing spatially inhomogeneous tissue to be modelled. As the results of these modifications can be compared directly with extracellular recordings, theoretical predictions can be tested even when spiking data is lacking. For example, spatial variations in synapse densities may be caused by the "patchy" projections made by excitatory neurons (Binzegger et al. 2007; Voges et al. 2010b; Bauer et al. 2012; Douglas and Martin 2004). Future research could incorporate the patchy projection model of (Voges et al. 2010a; Voges et al. 2010b) into our slice model to investigate how patchy connectivity affects network activity and resultant LFP across the slice.

We model the cortical layers as being flat, with boundaries at constant depths below the cortical surface. Neocortex is a folded structure, though, which is apparent even at the small scale of the slice – note the curved shaded regions in Fig. 9a showing the cortical surface and white matter boundaries, as well as the curved profile of gamma power across the MEA. Curves add further complications to the other inhomogeneities discussed above, in terms of neuronal densities, layer thicknesses and axonal arborisation variations. Additionally, the alignment of pyramidal apical dendrites is perpendicular to the cortical surface, so the alignment of the current dipoles arising from synaptic currents on pyramidal dendrites (Lindén et al. 2010; Nunez and Srinivasan 2006) varies across space, with implications for the measured LFP. VERTEX functions for specifying curved layer boundaries are currently under development so that future experiments can investigate the effects of curved surfaces on the measured LFP.

**Further considerations for LFP simulation**

In its current state, VERTEX is designed for investigating LFPs in medium to large-scale spiking neural networks, as these are most often used for modelling the activity of large neural populations. We have, therefore, only implemented simplified neuron models that do not include realistic active conductances that produce, for example, back-propagating dendritic spikes or sub-threshold membrane oscillations, which would also contribute to the LFP. As gamma oscillations are driven by synaptic interactions between populations, we consider this to be a reasonable simplification for our neocortical slice model. When investigating other dynamical regimes – such as sub-threshold oscillations in the absence of spiking (Hutcheon and Yarom 2000) – this simplification may not be appropriate. However, VERTEX will still be useful for investigating many research questions even with these simplifications. For example, most previous spiking neural network models use highly simplified neuron models, for which there is no general, reliable method for estimating the LFP (Einevoll et al. 2013). VERTEX allows researchers to implement similar networks using neuron models that produce a spatially realistic LFP, so that they can directly compare the LFPs produced by the spiking activity in their models to experimental data. Such comparisons may reveal both agreements and discrepancies between model and experiment, which might not have been apparent from comparisons of spiking alone. This was the case for our slice model: we could not directly compare spiking across space as it was massively under-sampled *in vitro*, but the simulated LFPs based on our prior knowledge of neuronal firing during gamma oscillations revealed that we can account for the observed phase inversion between L2/3 and L1, but cannot account for the spatial variation in gamma power with our current model. Future research to address this discrepancy is discussed above.

Several exciting experimental results have recently shown that neuronally generated electric fields impact on the membrane potentials of nearby neurons without requiring any synaptic contact. Such "ephaptic" coupling of neurons was investigated in models (Holt and Koch 1999) and, more recently, confirmed in experiments showing that such interactions could modulate oscillatory network activity (Fröhlich and McCormick 2010), entrain action potentials (Anastassiou et al. 2011) and potentially contribute to the spread of epileptiform activity (Zhang et al. 2014). We have purposefully ignored the contribution of ephaptic interactions in our model for the sake of simplicity, and have not incorporated the simulation of ephaptic coupling into the VERTEX simulator. While the results reported by Fröhlich and McCormick (2010) suggest that endogenous electric fields should be taken into account in models of oscillatory activity, they concentrated on neocortical slow oscillations, which are greater in amplitude than the gamma oscillations we modelled. However, the role of ephaptic interactions on network activity under different conditions must be investigated further. As VERTEX can simulate the LFP at arbitrary locations in a network, it would be possible to incorporate an ephaptic coupling mechanism that depended on the LFP. However, doing this rigorously would entail measuring the LFP near every compartment in the model, which is not feasible. Developing suitable approximation methods for incorporating realistic ephaptic coupling is therefore an important direction for future research. Similar methods could also be used for simulating artificially applied electric fields/currents, such as from extracellular stimulating electrodes.

Finally, the VERTEX simulator assumes a purely resistive, constant and homogeneous extracellular conductivity, with no frequency dependence (Pettersen et al. 2012). The extracellular medium's frequency filtering effects are not currently known for certain (Einevoll et al. 2013): some results have demonstrated an intrinsic low-pass filtering effect (Gabriel et al. 1996; Dehghani et al. 2010) potentially created by ionic diffusion (Bédard and Destexhe 2009), though direct measurements in macaque cortex *in vivo* found minimal frequency filtering from intrinsic tissue properties (Logothetis et al. 2007). If a frequency-dependent effect of the extracellular medium is confirmed by future studies, equations 1 and 2 (see Methods) can be modified to take this into account (Pettersen et al. 2012).

**Conclusion**

We have described the VERTEX simulation tool for simulating LFPs in large neuronal populations. VERTEX includes functionality for generating spatially constrained networks of several neuron populations, whose parameters are easily specified in Matlab structures. "Virtual electrodes" can be positioned at arbitrary locations in the model to simulate the LFP generated by the network. Parallel computing and code vectorisation, as well as the use of reduced compartmental neuron models, allows VERTEX to simulate network activity and LFPs in reasonable time. Finally, we simulated LFPs from a neocortical slice model and compared them with LFPs recorded from macaque neocortex *in vitro*, illustrating new avenues for research into spatial variations in the LFP signal. We hope that the VERTEX and our neocortical slice model will prove useful to other researchers investigating the relationship between neuronal circuit dynamics and experimental or clinical brain tissue recordings.

**METHODS**

**Software and simulation methods**

Spatial LFP characteristics of each individual compartmental neuron model were tested using LFPy as described in the Results section. LFPy simulations used a 0.125 ms time-step and Neuron's standard implicit Euler numerical integration method. Each simulation was run for 1250 ms simulation time, and the first 250ms were discarded to remove simulation start-up effects.

VERTEX is implemented in Matlab. It uses the Matlab Parallel Computing Toolbox for parallelisation, though it can also be run serially. Equations are integrated numerically using a second-order Runge-Kutta method (Press et al. 2007); we used a 0.03125 ms time-step unless otherwise specified. VERTEX incorporates the methods outlined in (Morrison et al., 2005) for parallel simulation, and the algorithms and data structures described in (Brette and Goodman 2011) for code vectorisation.

In both LFPy and VERTEX, extracellular potentials are calculated by summing the membrane currents of each compartment, weighted by distance from the electrode tips. The line-source method (Holt 1998), as used previously in (Lindén et al. 2010; Lindén et al. 2011; Pettersen and Einevoll 2008) is used to calculate the contribution from all dendritic compartments to the LFP, $\Phi_{dend}$, at a measurement point $r$ and time $t$:

$$\Phi_{dend}(r,t) = \sum_k \frac{I_{mem,k}(t)}{4\pi\sigma_{ex}\Delta s_k} \log \left| \frac{\sqrt{h_k^2 + \rho_k^2} - h_k}{\sqrt{l_k^2 + \rho_k^2} - l_k} \right|, \qquad (1)$$

(Holt 1998), where $I_{mem,k}$ is the membrane current from compartment $k$, $\sigma_{ex}$ is the extracellular conductivity, $\Delta s_k$ is the length of compartment $k$, $\rho_k$ is the perpendicular distance from compartment $k$, $h_k$ is the longitudinal distance from the end of compartment $k$, and $l_k = \Delta s_k + h_k$ is the longitudinal distance from the start of the compartment. As in (Lindén et al. 2010; Lindén et al. 2011; Pettersen and Einevoll 2008), somatic compartments are modelled as point current sources in VERTEX:

$$\Phi_{soma}(r,t) = \sum_s \frac{I_{mem,s}(t)}{4\pi\sigma_{ex}r_s}, \qquad (2)$$

(Nunez and Srinivasan 2006), where $r_s$ is the distance between point $r$ and the centre of soma $s$. The total extracellular potential measurement at point $r$ is then $\Phi = \Phi_{soma} + \Phi_{dend}$. In all simulations, we used a value of $\sigma_{ex} = 0.3$ S/m (Hämäläinen et al., 1993).

LFPy simulations to test the model reduction method (Bush and Sejnowski 1993) were run on an Intel Core-i7 based PC running Ubuntu Linux 11.10 using a pre-release version of LFPy, NEURON 7.1 and Python 2.7.2, and an Intel Xeon E5640 workstation running Linux Mint 16 using LFPy 1.0 with NEURON 7.3 and Python 2.7.5. The LFPy vs. VERTEX performance comparison was run on the same Intel Xeon E5640 workstation, using Matlab 2013a. All other VERTEX simulations were run on a 48-core HP ProLiant server running CentOS Linux 5.8 with Matlab R2012b. Parallel simulations of were run on 12 cores unless otherwise specified. The code, as well as documentation and tutorials, will be made available at http://www.biological-networks.org/ upon publication.

**Neocortical slice model**
The neocortical slice model contained fifteen neuron populations, defined by location, connectivity, morphology, dynamics, and type of neurotransmitter effect (excitatory or inhibitory). We used the naming convention from (Binzegger et al. 2004) as adapted in (Izhikevich and Edelman 2008), defining the groups listed in Table 1 (full model parameters are given in Tables ESM1-ESM9). Individual neurons are represented by compartmental models with 7, 8 or 9 compartments, derived from the neuron models given in (Mainen and Sejnowski 1996) using the compartmental reduction method of (Bush and Sejnowski 1993). Compartmental structure and neuron parameters are given in Fig. ESM2, Table ESM1 and Table ESM2. Our connectivity data is from cat visual cortex (Binzegger et al. 2004), so we took parameters for the neuronal density and layer boundaries from the same source. We scaled the layer boundaries to increase the total cortical depth to 2.6 mm, which was approximately the cortical depth in the comparison experimental slice (established by post hoc histology, not shown).

VERTEX is designed for specifying models in 3D space, giving all neuronal compartments 3D start and end coordinates. For the neocortical slice, we defined the $z$-axis to be the cortical depth from white matter through the layers to the cortical surface, with the border between layer 6 and the white matter set to $z = 0$ mm. The $x$- and $y$-axes ran parallel to the cortical surface, with the $y$-axis pointing along the thickness of the slice, and the $x$-axis along the slice width. The boundaries between cortical layers were then defined as $x$-$y$ planes with constant depth $z_l$. Layer 1 was a-neuronal, and layers 2 and 3 were combined. The total model size was then specified by the cortical depth $z_{max}$, the thickness of the slice $y_{max}$, the width of the slice $x_{max}$, and the neuronal density $D$, with the total number of neurons calculated as $N = x_{max} \times y_{max} \times z_{max} \times D$. The model slice had dimensions $x_{max} = 4.4$ mm, $y_{max} = 0.4$ mm and $z_{max} = 2.6$ mm, and $D = 38\ 335$ neurons/mm$^3$, resulting in a model size of 175 421 neurons. We then positioned neurons by placing their somas at random $x$, $y$ and $z$ values constrained by $x_{max}$, $y_{max}$ and the $z_l$ boundaries of the containing layer, and rotating them by random angles. Pyramidal cells had their apical dendrites aligned parallel to the $z$-axis.

All neurons could form connections within their group and with neurons from all other neuron groups, according to the values given in Table ESM3. Connections were also constrained by the axonal arborisation radii of the presynaptic neurons, taken from (Blasdel et al. 1985; Fitzpatrick et al. 1985) as adapted in (Izhikevich and Edelman 2008). Arborisations were considered in 2D: on the $x$-$y$ plane on a per-layer basis. We assumed an isotropic Gaussian spatial distribution of connections centred on the presynaptic neuron (Hellwig 2000), setting the arborisation radius equal to two standard deviations of the Gaussian kernel, so that ~91% of connections were contained within the specified arborisation radius. Arborisation radii are given in Table ESM4. When deciding on the targets of a pre-synaptic neuron $i$ in layer $l$, we calculated the expected number of connections made by $i$ in $l$ remaining inside the slice by multiplying the number of connections specified in Table ESM3 by the ratio $\zeta_{li}$, defined as the integral of the kernel within the slice boundaries:

$$\zeta_{li} = \int_{a_{yi}}^{b_{yi}} \int_{a_{xi}}^{b_{xi}} \frac{1}{2\pi\sigma_{li}^2} \exp\left[-\left(\frac{x^2 + y^2}{2\sigma_{li}^2}\right)\right] dx\,dy \qquad (3)$$

$$= \frac{1}{4}\left\{\left[\mathrm{erf}\left(\frac{a_{xi}}{\sqrt{2}\sigma_{li}}\right) - \mathrm{erf}\left(\frac{b_{xi}}{\sqrt{2}\sigma_{li}}\right)\right] \times \left[\mathrm{erf}\left(\frac{a_{yi}}{\sqrt{2}\sigma_{li}}\right) - \mathrm{erf}\left(\frac{b_{yi}}{\sqrt{2}\sigma_{li}}\right)\right]\right\},$$

where $a_{xi}$ is the distance from neuron $i$ to the left edge of the slice, $b_{xi}$ the distance to the right edge, $a_{yi}$ the distance to the front edge, $b_{yi}$ the distance to the back edge, $\sigma_{li}$ is half the arborisation radius of $i$ in layer $l$ (see Fig. 5), and erf the Gaussian error function (this solution is valid provided that $a_{xi}$ and $a_{yi}$ are negative, and $b_{xi}$ and $b_{yi}$ are positive). Pyramidal neuron dendrites span several layers above their soma layer, so we used the connectivity statistics provided per layer for pyramidal neurons in (Binzegger et al. 2004). Axonal transmission delays were calculated as the Euclidean distance between the presynaptic and postsynaptic neurons' somas divided by the axonal transmission speed of 0.3 m/s (Hirsch and Gilbert, 1991), plus a constant synaptic delay of 0.5 ms to account for the time taken for neurotransmitter release and binding (Katz and Miledi, 1965). All these modelling decisions are handled by the initialisation functions in VERTEX, which also allow models to be initialised with uniform spatial connectivity profiles, no synapse reduction, arbitrary delay times, and in a cylindrical shape rather than the cuboid of our slice model.

Synapse weights are specified in Table ESM5. We included AMPA and GABA$_A$ type conductance-based synapses (the minimal set of synapse types required for generating gamma). When a neuron fired a spike, the synaptic conductance at the contacted target compartments increased by the relevant synaptic weight after the relevant axonal delay time, then decayed exponentially. VERTEX currently includes current-based and conductance-based models of single-exponential and alpha synapses.

We stimulated our model to mimic the bath application of kainate, which excites the pyramidal axonal plexus, providing the neurons with excitatory drive. We simulated this by applying independent random input currents to each neuron, modelled as Ornstein-Uhlenbeck processes [similar to (Arsiero et al. 2007)]. Input current parameters are given in Table ESM6. VERTEX can provide random inputs to neurons as either currents or membrane conductance fluctuations.

### *In vitro* experimental methods
All experiments were carried out in accordance with the European Communities Council Directive 1986 (86/609/EEC), the US National Institutes of Health Guidelines for the Care and Use of Animals for Experimental Procedures, and the UK Animals Scientific Procedures Act.

### Surgical preparation
The monkey (*Macaca mulatta*, male, 8 years old) used in this study was subject to experiments *in vivo* involving extracellular recording of neural activity and local drug application (iontophoresis). All tissue samples used in this study were taken from intact brain areas that were not the subject of studies performed before tissue extraction. Extraction was performed under general anaesthesia, which was maintained over the course of four days. For the anaesthesia the animal was initially sedated with a 0.1 ml/kg ketamine intra-muscular injection (100mg/ml). Thereafter, bolus injections of propofol were administered intravenously to allow for tracheotomy and placement of catheters for measuring intra-arterial and central venous blood pressure. During surgery, anaesthesia was maintained by gaseous anaesthetic (2.5-3.9% sevoflurane) combined with continuous intravenous application of an opioid analgesic (Alfentanil, 120µg/kg/h), a glucocorticoid (Methylprednisolone, 5.4mg/kg/h) and saline (50ml/h). The animal's rectal temperature, heart rate, blood oxygenation and expired $CO_2$ were monitored continuously during anaesthesia.

### Slice preparation
Macaque neocortical samples were routinely obtained from the inferior temporal gyrus. This was confirmed by post-hoc anatomical examination of the fixed (paraformaldehyde) whole brain. Following resection, cortical samples were immediately placed in ice-cold sucrose artificial cerebrospinal fluid (ACSF) containing: 252 mM sucrose, 3 mM KCl, 1.25 mM $NaH_2PO_4$, 2 mM $MgSO_4$, 2 mM $CaCl_2$, 24 mM $NaHCO_3$, and 10 mM glucose. Neocortical slices containing all layers were cut at 450µm (Microm HM 650V), incubated at room temperature for 20-30 min, then transferred to a standard interface recording chamber at 34-36°C perfused with oxygenated ACSF containing: 126 mM NaCl, 3 mM KCl, 1.25 mM $NaH_2PO_4$, 1 mM $MgSO_4$, 1.2 mM $CaCl_2$, 24 mM $NaHCO_3$, and 10 mM glucose. Persistent gamma frequency oscillations were induced by the application of kainate (400-800nM) to the circulating ACSF and were deemed stable if there was no change to frequency or power after 1 hour. In general we did not observe spontaneous network activity in the slices before the bath addition of kainate. LFP recordings were taken using multichannel 10x10 silicon electrodes with an inter-electrode distance of 400µm (Utah array, Blackrock Microsystems, Salt Lake City, UT, USA). Time series were digitally sampled at 10 kHz.

### Data processing and analysis
Data processing and analysis was performed in Matlab R2012b. We used the same processing chain for both simulated and experimental recordings, except that common average re-referencing, line noise removal and renormalisation were only applied to the experimental recordings. For LFP analysis, recordings were first re-referenced to the common average, then resampled at 1 kHz. We removed line-noise and harmonics by band-pass filtering each recording at 49-51 Hz, 99-101 Hz, 149-151 Hz, 199-201 Hz and 249-251 Hz (symmetrical Butterworth filter, $8^{th}$ order) and subtracting the resulting signal from the original signal. The recordings were then band-pass filtered between 2 Hz and 300 Hz (symmetrical FIR filter, Kaiser window, $2000^{th}$ order). We restricted our analysis to an 18 second segment of the recording that was identified as artefact-free in all channels by visual inspection of the filtered traces. After filtering, these segments were normalised to zero mean, unit standard deviation to facilitate signal comparison across the MEA.

Power spectra were calculated using the Thomson multitaper method with a time-bandwidth product of 10 (19 tapers) for experimental recordings and 3 (5 tapers) for the shorter simulated recordings, with estimated 95% confidence intervals calculated using a chi-squared approach. Total gamma power at each electrode was calculated by taking the integral of the power spectrum between 20 Hz and 40 Hz. Gamma power between electrodes was estimated by bicubic interpolation between electrode locations.

**ACKNOWLEDGEMENTS**

We thank the anonymous reviewers for their constructive and insightful comments, which helped us to greatly improve the final manuscript. RJT was supported by the BBSRC (BB/F016980/1). MA and MK were funded as part of the CARMEN e-science project (http://www.carmen.org.uk) by the EPSRC (EP/E002331/1). MK also received support from the EPSRC (EP/G03950X/1, EP/K026992/1). AT was supported by the MRC, BBSRC and the Wellcome Trust. MOC was supported by the EPSRC, MRC and the Wolfson Foundation. MW is supported by a Wellcome Trust Senior Investigator Award.

**Table 1** Neuron groups, abbreviations, and number of compartments within our model. Basket interneurons are in L2/3, L4, L5 and L6. Non-basket interneurons are in L2/3, L4 and L5. Compartmental structures are shown in Fig. ESM2

| Abbreviation | Neuron group description | Proportion of total model [%] | Compartments |
|---|---|---|---|
| P2/3 | pyramidal neurons in layer 2/3 (L2/3) | 27.4 | 8 |
| SS4(L4) | spiny stellate neurons in L4 projecting to L4 | 9.7 | 7 |
| SS4(L2/3) | spiny stellate neurons in L4 projecting to L2/3 | 9.7 | 7 |
| P4 | pyramidal neurons in L4 | 9.7 | 8 |
| P5(L2/3) | pyramidal neurons in L5 projecting to L2/3 | 5.0 | 9 |
| P5(L56) | pyramidal neurons in L5 projecting to L56 | 1.4 | 9 |
| P6(L4) | pyramidal neurons in L6 projecting to L4 | 14.1 | 9 |
| P6(L56) | pyramidal neurons in L6 projecting to L56 | 4.7 | 9 |
| B# | basket interneurons in L# | 13.7* | 7 |
| NB# | non-basket interneuron in L# | 4.7* | 7 |

*Proportions given for the whole model rather than per layer; proportions per layer are given in Table ESM3.

**Fig. 1** Comparison of simulated LFPs from the Bush and Mainen cell models. Top (red): L2/3 pyramidal neuron, middle (green): spiny stellate cell (morphology also used for interneurons), bottom (blue): L5 pyramidal neuron. **a** Comparison of original and reduced multi-compartment models of each neuron type. **b** Range and magnitude of simulated LFPs. Circles show values for the original cell reconstruction populations, triangles for the reduced neuron model populations. Light red dashed lines in the top panel and light blue circles in bottom panel show values for the extra cat pyramidal neurons tested, as described in the main text. All *y*-axis values in μm. **c** Overlap of the 95% confidence intervals for the estimated LFP power spectra produced by each population in each layer shaded dark. Non-overlapping sections of the 95% confidence intervals are shaded light. Power is plotted in dimensionless, normalised units for ease of comparison

**Fig. 2** Comparison of simulated LFPs from the Bush and Mainen cell models for highly correlated input at the soma compartment. Top (red): L2/3 pyramidal neurons, bottom (blue): L5 pyramidal neurons. **a** Range and magnitude of simulated LFPs. Bright red/blue lines show range and magnitude values for the Mainen cell populations, dark red/blue lines show range and magnitude values for the Bush cell populations. The faded red/blue dashed lines show these values for the additionally tested cell populations in L2/3 and in L5. Grey dashed lines show layer boundaries, black solid lines show the maximum and minimum soma depths. All $y$-axis values in μm. **c** Overlap of the 95% confidence intervals for the estimated LFP power spectra produced by the L2/3 and L5 pyramidal neuron populations at each electrode location shaded dark (correlated input at soma). Non-overlapping sections of the 95% confidence intervals are shaded light. Power is plotted in dimensionless, normalised units for ease of comparison. Comparisons for only 13 out of the 26 LFP measurement points for the L5 populations are shown for ease of visualisation

**Fig. 3** Overview of the VERTEX simulation software. **a** Simulation workflow. The user provides parameters as Matlab structures to setup the neuron populations, position them in layers, connect them together, and simulate their dynamics and the resultant LFPs. Functionality to export to NeuroML is currently under development. **b** Example program structure. The main simulation program only requires calls to the initialiseNetwork() function and the runSimulation() function, with the information required to setup the simulation specified in separate script files

**Fig. 4** Parallel simulation performance with increasing numbers of Matlab workers (i.e. parallel processes). Top row: model initialisation times for **a** the 9 881 neuron model and **b** the 123 517 neuron model. Bottom: simulation times for 1 second of biological time for **c** the 9 881 neuron model and **d** the 123 517 neuron model. Thick black lines indicate linear speed scaling; legends indicate the number of electrodes used in each simulation run. The sub-linear speed-up in the small model is due to the decreasing relative performance influence of code vectorisation for smaller matrices (see Results)

**Fig. 5** Slice model structure and individual neuron dynamics. **a** Layer boundaries are given in μm. Subsets of soma locations from each neuron group are shown in faded black for excitatory neurons, or faded magenta for inhibitory neurons. Triangles represent pyramidal neuron somas, stars are spiny stellate cell somas, circles are basket interneuron somas and diamonds non-basket interneuron somas. One example full cell is shown for each neuron group, in solid black for excitatory neurons or solid magenta for inhibitory neurons. Compartment lengths are to scale; compartment diameters are not. Black circles are virtual electrode positions (first 8 rows shown). **b** Responses to step-current injections into the soma compartment of each neuron type. Spikes are detected and cut-off at $V_t + 5$ mV; we extend the spike trace up to +10 mV for illustrative purposes. Step current magnitudes were 0.5 nA for the P2/3 neuron, 0.333 nA for the SS neuron, 1.0 nA for the P5 neuron, 0.75 nA for the P6 neuron, and 0.4 nA for the B and NB interneurons

**Fig. 6** Changes in connectivity between neuron groups after slice cutting. **a** Expected number of connections from population of presynaptic neurons (columns) onto single postsynaptic neurons (rows) before slicing, based on the data from (Binzegger et al. 2004) **b** Illustration of the effect of slice cutting on a presynaptic neuron's (light green dot) axonal arborisation (shaded area). Figure orientation is as if looking down onto the surface of the brain, with slice boundaries indicated by the black bounding box. Connections within the green shaded area remain, but those in the grey shaded areas are removed by slicing. **c** Connectivity in the cortical slice model, as altered from **a** by slice cutting. While overall connection number decreases (note different scale bars), some connections are affected more than others because of differing axonal arborisation sizes

**Fig. 7** Spike raster and individual neuron responses during gamma oscillation. **a** Spike raster showing spiking activity of 5% of all the neurons in the model (reduced number shown for clarity). Boundaries between neuron groups marked in cyan. Note strong persistent gamma oscillation in L2/3, with weaker oscillation in L5. **b** Example soma membrane potential plots for the various neuron types. Most neurons fire sparsely, while B2/3 and B5 neurons fire on most oscillation periods. Note occasional spike doublet firing in the B2/3 neuron. Spikes are cut-off at $V_t + 5$ mV in the simulation; we extend them up to 10 mV here for illustrative purposes. **c** Close-up of P2/3 neuron soma membrane potential (cut-off -45 mV). Scale-bar: 5 mV

**Fig. 8** Illustration of the gamma oscillation mechanism in the model. **a** Spike raster of 250 ms from a simulation of a model with the same parameters as that shown in Fig. 6. For clarity, spikes form only 5% of the neurons are shown. A gamma oscillation is apparent in layers 2/3 and 5. **b** Zoomed spike raster showing only neurons in layer 2/3. Spikes from only 1% of the neurons are shown. **c** LFP recording from the virtual electrode with the highest gamma power in the LFP. **d** Power spectrum of the LFP from this electrode, calculated for 1.5 s simulation time, showing a clear gamma peak. **e-h** Same as **a-d**, but with synaptic weights from P2/3 cells to B2/3 cells reduced to 1% of their original value. **e-f** show B2/3 cell firing is greatly reduced, as they are not receiving excitation from the P2/3 cells. No gamma oscillation emerges. **i-l** same as **a-d**, but with synaptic weights from B2/3 cells to P2/3 cells reduced to 1% of their original value. B2/3 cells fire rapidly and randomly: they are driven by the P2/3 cells but they cannot synchronise them as their synapses are too weak. No gamma oscillation emerges. **m-p** same as **a-d**, but with the mean and standard deviation of the stochastic input current to the B2/3 cells increased by 50%. P2/3 cell firing is suppressed by the increased B2/3 cell firing, so no gamma oscillation occurs.

**Fig. 9** Comparison of experimental (**a-c**) and simulated (**d-f**) MEA recordings. **a** Map of gamma frequency power across the electrode array *in vitro*. Electrode positions shown as grey dots, corner numbers indicate electrode IDs. Shaded areas show where electrodes were discounted because they fell either outside the slice boundaries or within the white matter. Gamma power is strongest at the top of the slice, corresponding to L2/3. **b** Example experimental LFP traces from electrodes 41 to 44 [indicated by grey rectangle in **a**]. Traces have been normalised to unit standard deviation for ease of comparison. **c** Cross correlation of signals from electrodes 41 to 44 with signal from electrode 42, illustrating phase inversion in the signal from electrode 41. This electrode was identified as being in layer 1 by post hoc histology (not shown). Gamma map & cross-correlations estimated from 18 s of data. **d-f** As **a-c**, but for the neocortical slice model (gamma map & cross-correlations estimated from 1.5 s of simulation data)

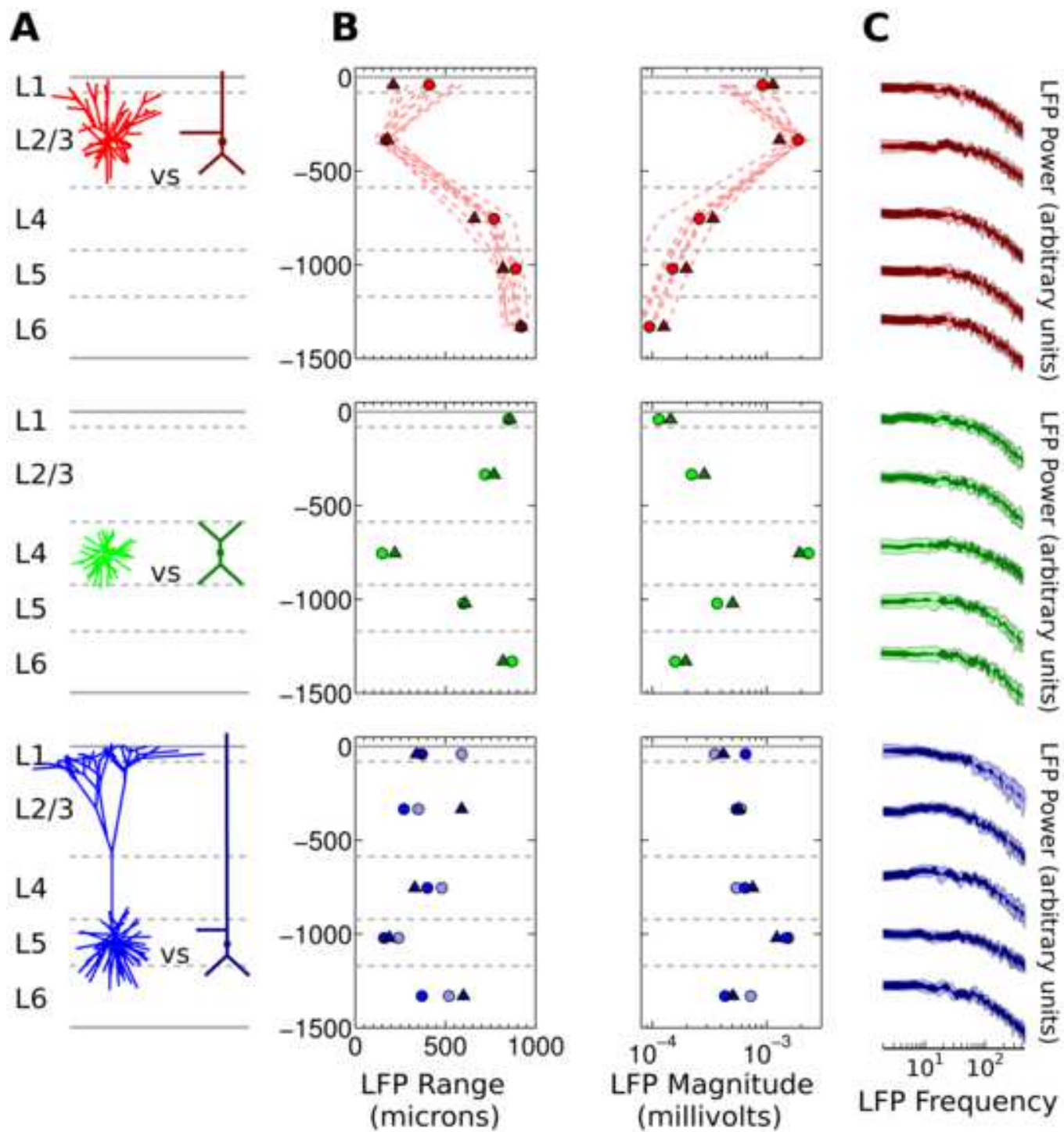

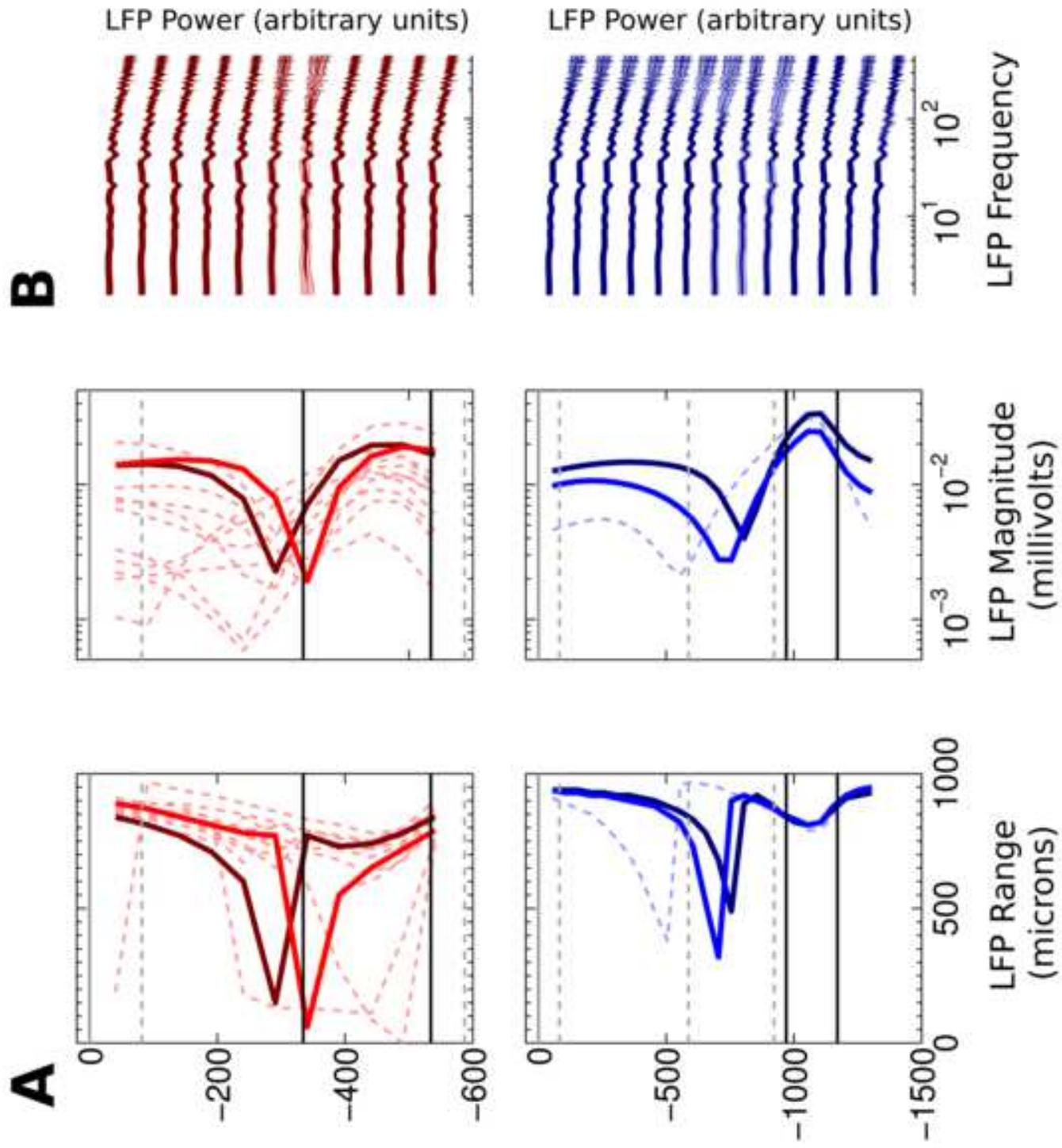

Fig 2
Click here to download high resolution image

Fig 3
Click here to download high resolution image

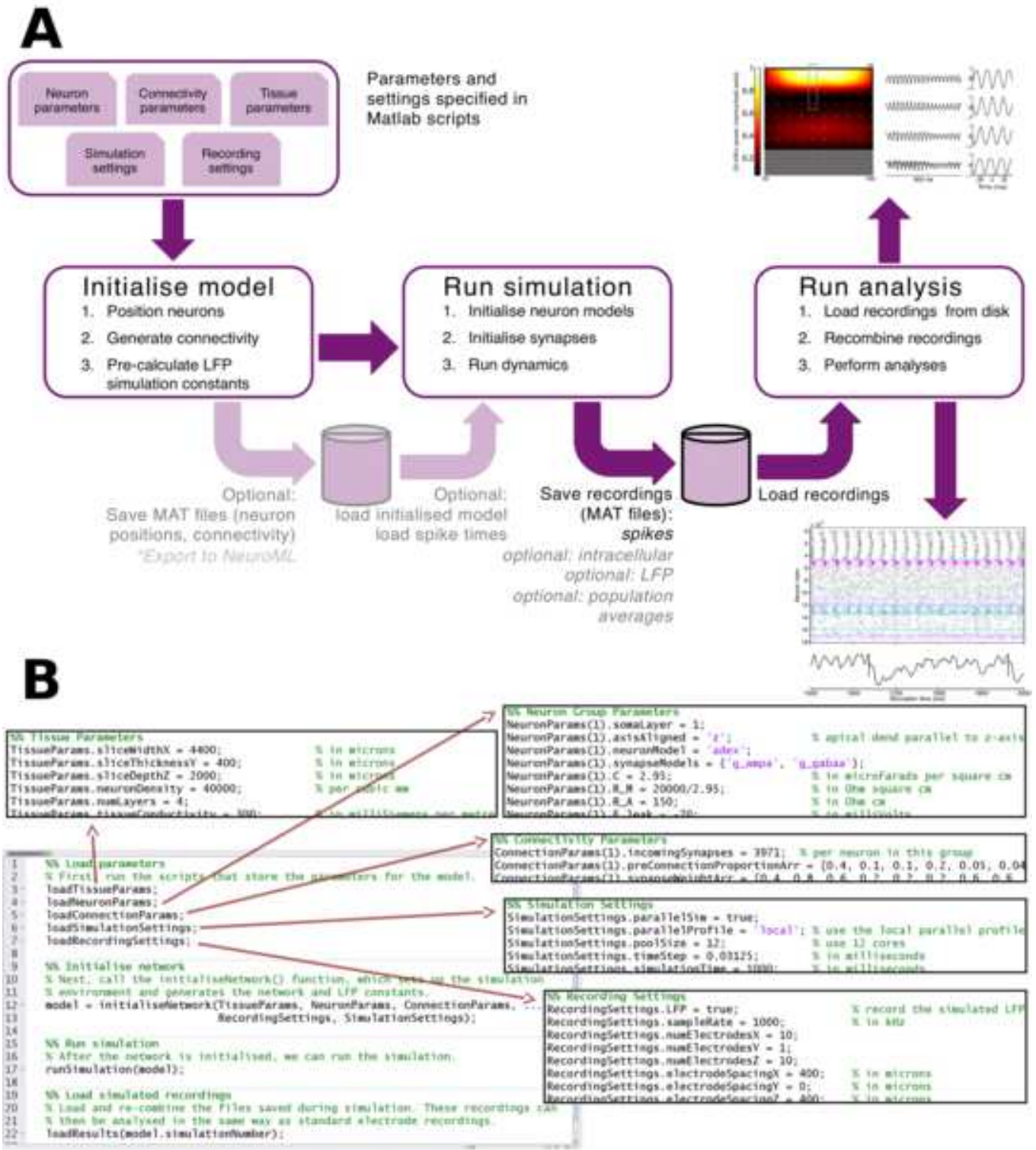

Fig 4
Click here to download high resolution image

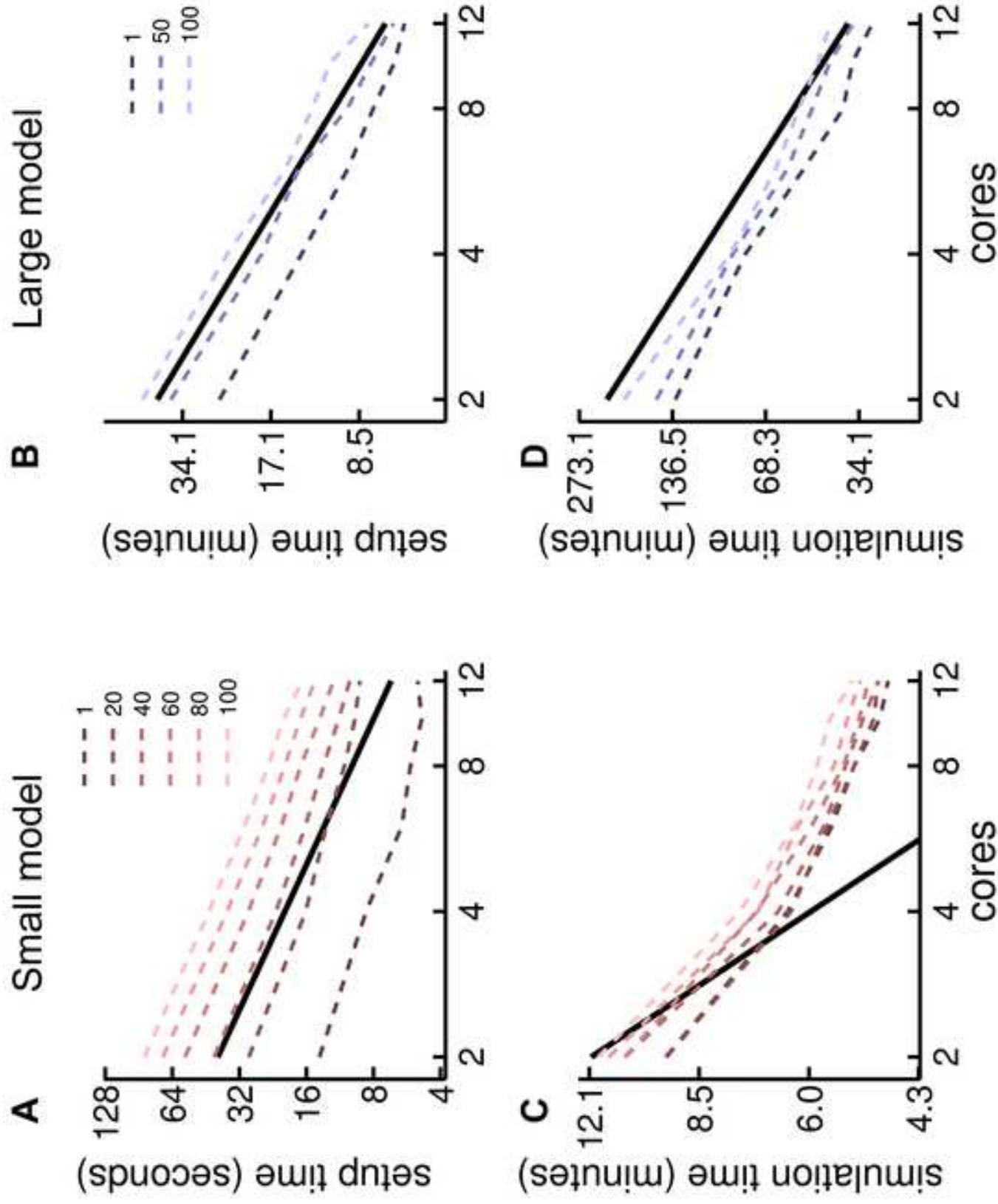

Fig 5
Click here to download high resolution image

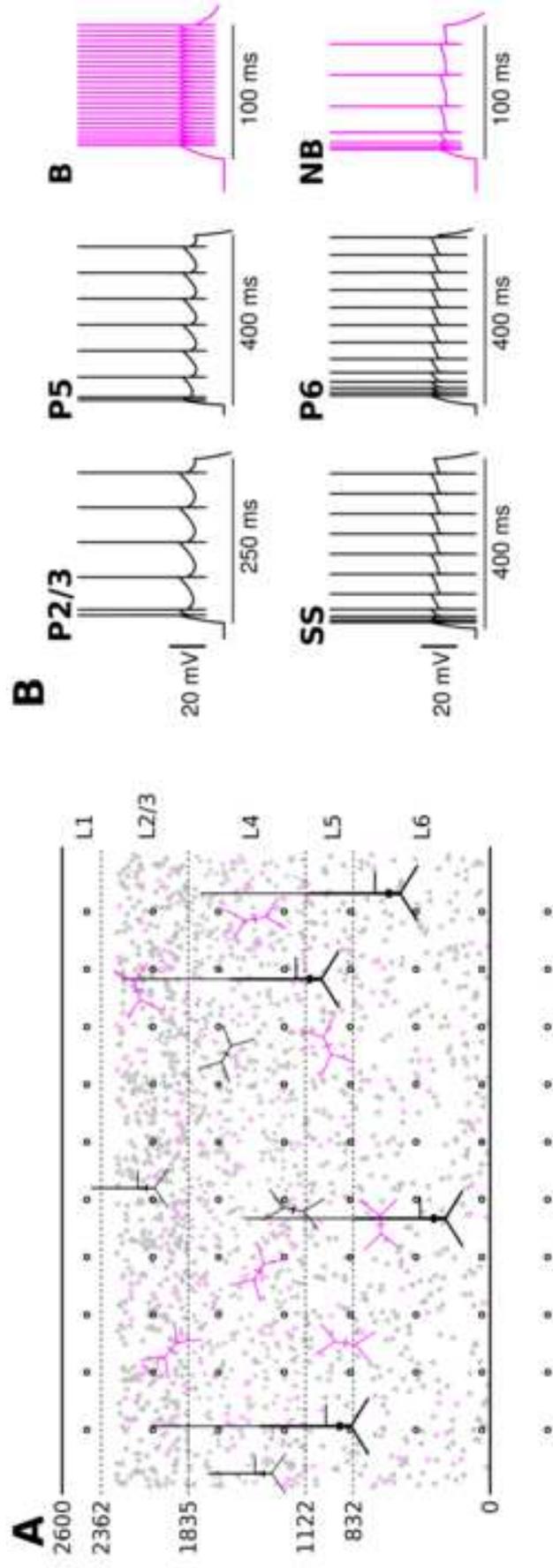

Fig 6
Click here to download high resolution image

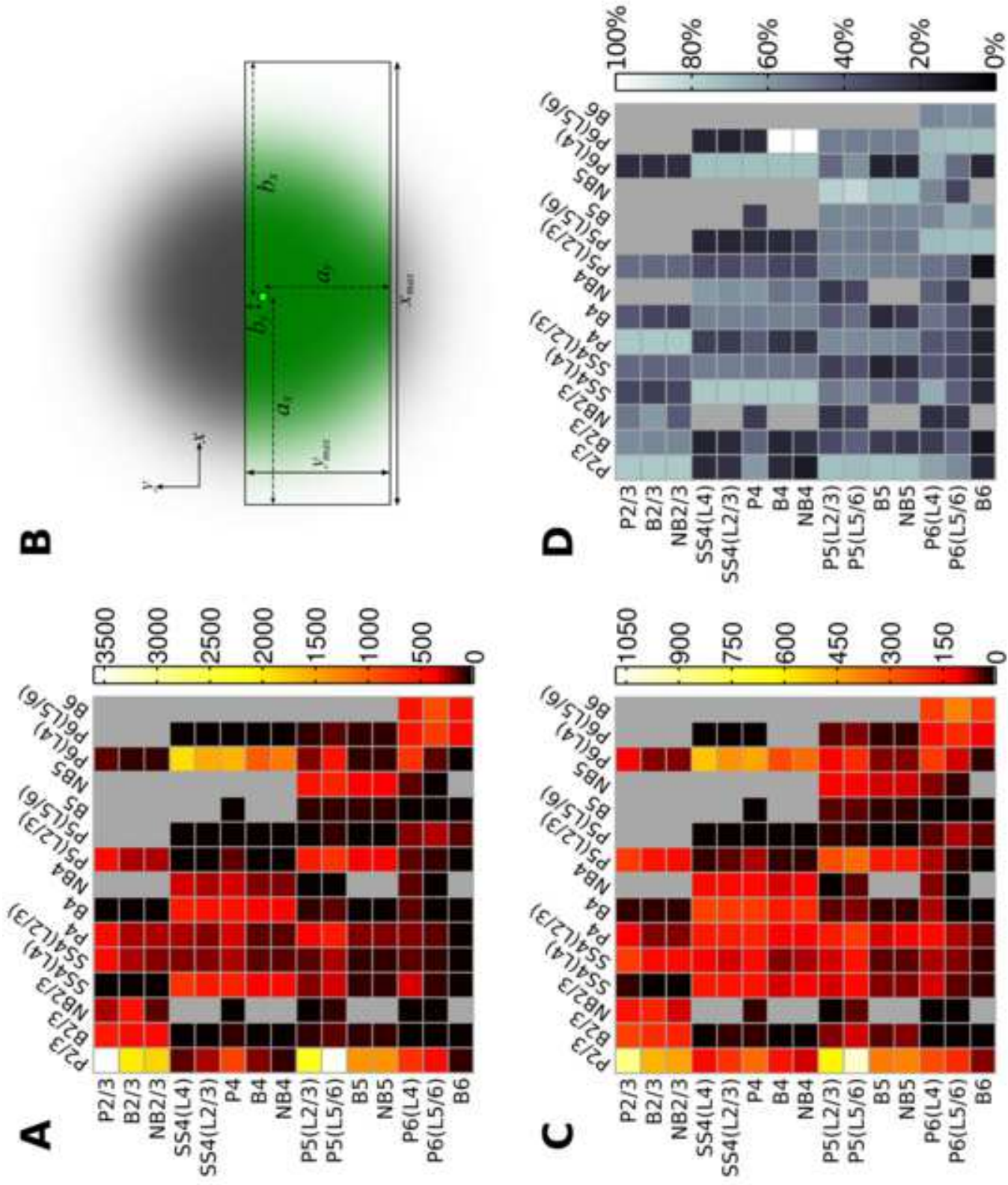

Fig 7
Click here to download high resolution image

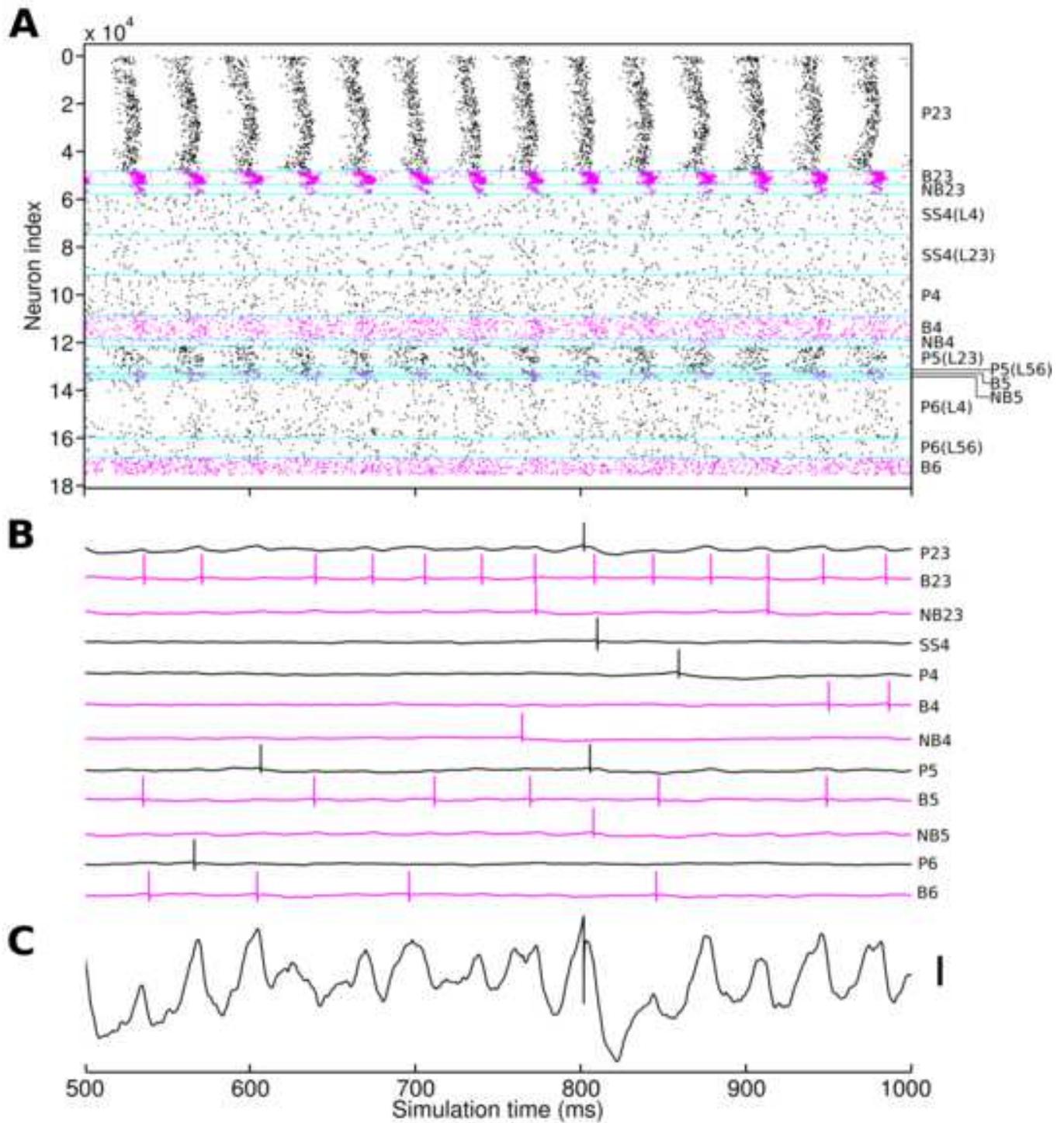

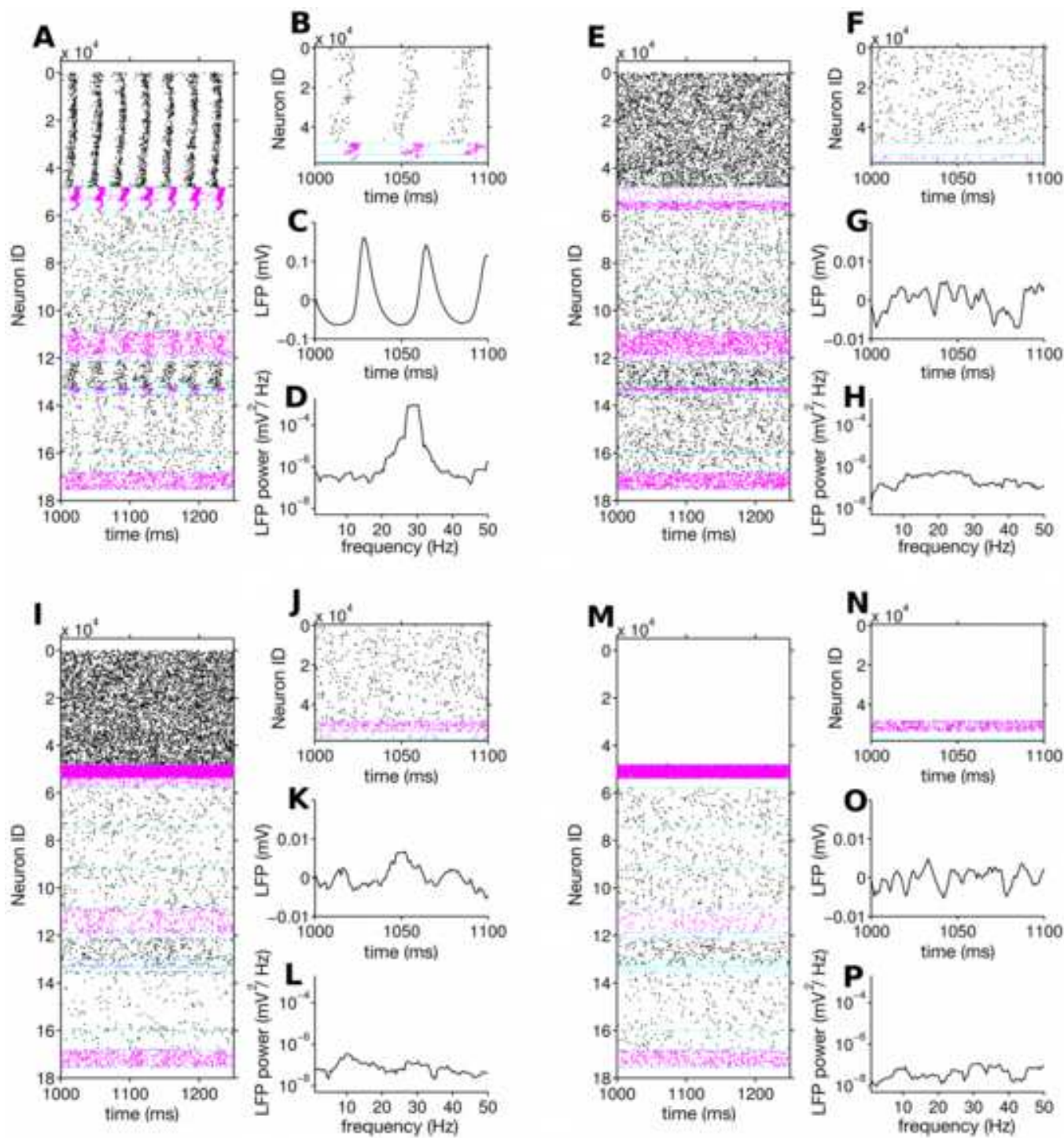

Fig 9
Click here to download high resolution image

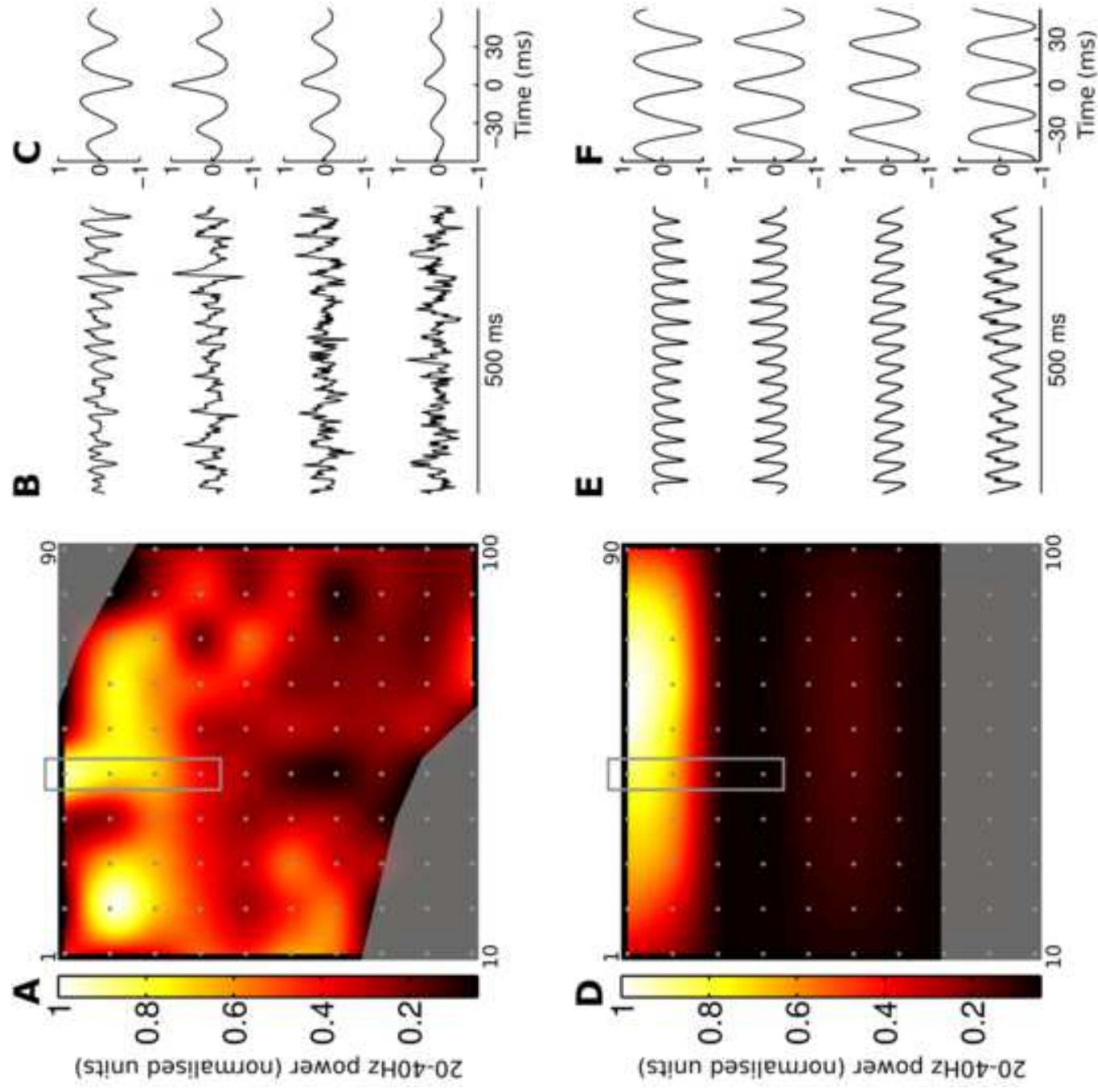

ELECTRONIC SUPPLEMENTARY MATERIAL

**Virtual Electrode Recording Tool for EXtracellular potentials (VERTEX): Comparing multi-electrode recordings from simulated and biological mammalian cortical tissue**

**Brain Structure and Function**

Richard J Tomsett, Matt Ainsworth, Alexander Thiele, Mehdi Sanayei, Xing Chen, Alwin Gieselmann, Miles A Whittington, Mark O Cunningham* and Marcus Kaiser*

*Joint senior authors

**Correspondence:**
Dr Marcus Kaiser
School of Computing Science
Newcastle University
Claremont Tower,
Newcastle upon Tyne, NE1 7RU, United Kingdom

Telephone: +44 191 208 8161
Fax: +44 191 208 8232
Email: m.kaiser@ncl.ac.uk

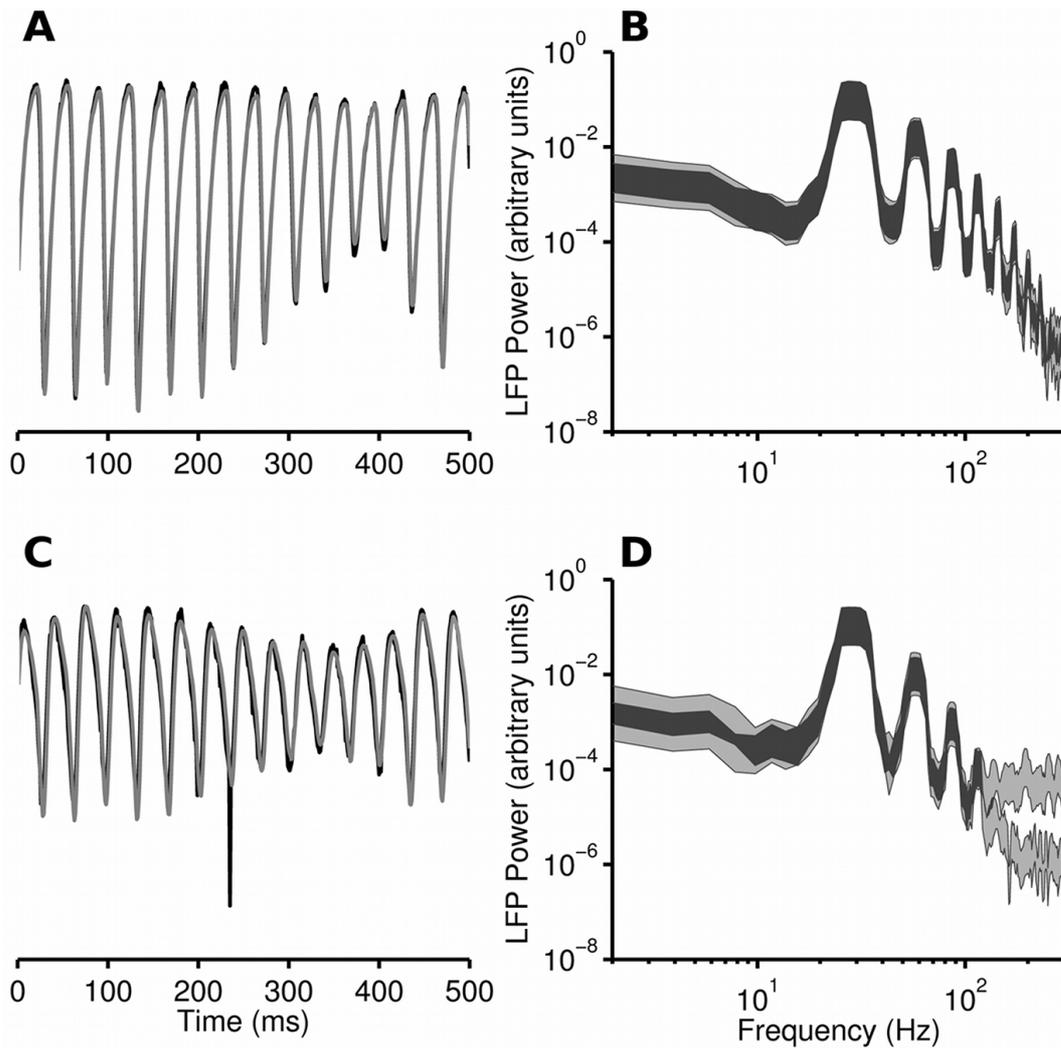

**Fig. ESM1** Comparison of simulated LFPs when using purely passive neurons with imported spike times, and when using the AdEx spiking model in the neocortical slice model (related to Results: Spike Import section). Traces in **a** and **c** have been normalised to zero mean, unit standard deviation so that the shape of the LFP signal can be more easily compared, but are otherwise unfiltered. The adaptive current of the AdEx mechanism introduces an offset in the simulated LFPs, but does not dramatically affect the shape of the signal. **a** Simulated LFPs from electrode 42; AdEx version in black, passive version in grey. **b** Power spectral density overlap of these signals (estimated for 500ms signal), with overlapping parts of the estimated spectrum shown in dark grey and non-overlapping shown in light grey. **c** As a, but for electrode 97. Note the small, sharp spike at ~230ms in the AdEx signal. This is a result of the AdEx reset mechanism creating a very short, fast current from a neuron very close to the electrode. **d** As **b**, but for electrode 97. Power spectra diverge above ~100 Hz due to high frequency spike contamination in the AdEx model, but match closely below 100 Hz. We recommend that, while the LFP can be estimated directly when using AdEx spiking networks, results should be checked by using VERTEX's spike import function if a true representation of the synaptic contribution to the LFP is sought at higher frequencies

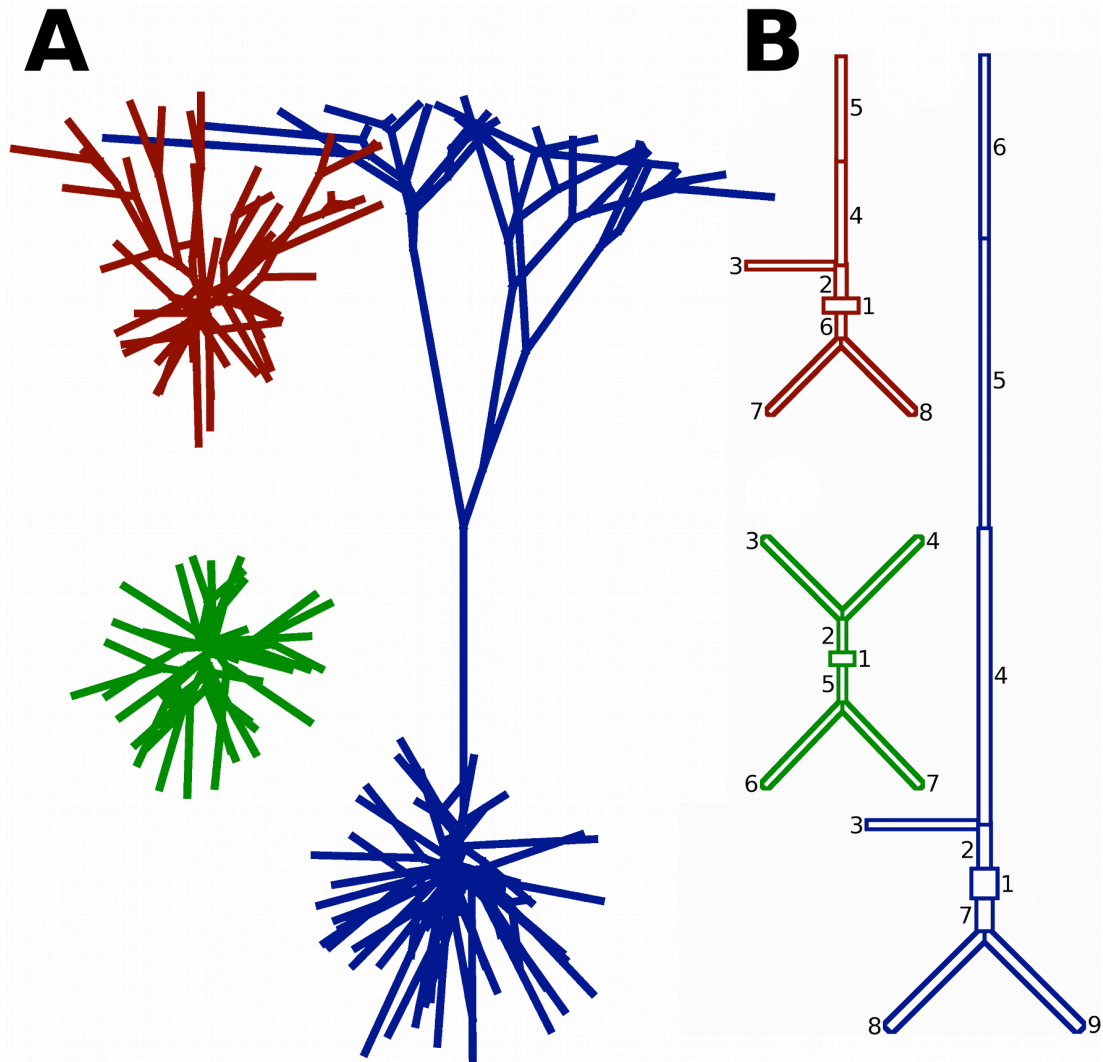

**Fig. ESM2 a** Compartmental structures of morphological cell reconstructions from (Mainen and Sejnowski 1996). The layer 2/3 pyramidal cell, layer 4 spiny stellate cell and layer 5 pyramidal cell are shown in red, green and blue, respectively. **b** Compartmental models reduced from the structures in **a** according to the method in (Bush and Sejnowski, 1993). Compartment numbers correspond to those in Table ESM1

**Table ESM1** Reduced neuron model compartment dimensions, used in all reported simulations

| Compartment number | P2/3, P4 | | P5, P6 | | SS, B, NB | |
|---|---|---|---|---|---|---|
| | Length (μm) | Diameter (μm) | Length (μm) | Diameter (μm) | Length (μm) | Diameter (μm) |
| 1 | 13 | 29.80 | 35 | 25.00 | 10 | 24.00 |
| 2 | 48 | 3.75 | 65 | 4.36 | 56 | 1.93 |
| 3 | 124 | 1.91 | 152 | 2.65 | 151 | 1.95 |
| 4 | 145 | 2.81 | 398 | 4.10 | 151 | 1.95 |
| 5 | 137 | 2.69 | 402 | 2.25 | 56 | 1.93 |
| 6 | 40 | 2.62 | 252 | 2.40 | 151 | 1.95 |
| 7 | 143 | 1.69 | 52 | 5.94 | 151 | 1.95 |
| 8 | 143 | 1.69 | 186 | 3.45 | - | - |
| 9 | - | - | 186 | 3.45 | - | - |

**Table ESM2** Neuron model parameters, used in all reported simulations (simulations of purely passive neurons only have $C_m$, $R_m$, $R_a$ and $E_l$ specified)

| Neuron type | $C_m$ (μFcm$^{-2}$) | $R_m$ (kΩcm$^2$) | $R_a$ (Ωcm) | $E_l$ (mV) | $V_T$ (mV) | $\Delta_T$ (mV) | $\alpha$ (nS) | $\tau_w$ (ms) | $\beta$ (pA) | $v_{reset}$ (mV) |
|---|---|---|---|---|---|---|---|---|---|---|
| **P2/3, P4** | 2.96 | 6.76 | 150 | -70 | -50 | 2.0 | 2.60 | 65 | 220 | -60 |
| **SS4** | 2.95 | 5.12 | 150 | -70 | -50 | 2.2 | 0.35 | 150 | 40 | -70 |
| **P5** | 2.95 | 6.78 | 150 | -70 | -52 | 2.0 | 10.00 | 75 | 345 | -62 |
| **P6** | 2.95 | 6.78 | 150 | -70 | -50 | 2.0 | 0.35 | 160 | 60 | -60 |
| **B** | 2.93 | 5.12 | 150 | -70 | -50 | 2.0 | 0.04 | 10 | 40 | -65 |
| **NB** | 2.93 | 5.12 | 150 | -70 | -55 | 2.2 | 0.04 | 75 | 75 | -62 |

**Table ESM3** Model composition. Neuron population sizes are given as percentage of total model size. The maximum number of synapses received by a postsynaptic neuron is specified per-layer for pyramidal neurons, whose apical dendrites span several layers. The proportions of these synapses made by each presynaptic neuron group are given in percentages of these maximal synapse numbers. Neurons in the slice model receive fewer than the maximum number of possible synapses because of the effects of slice cutting (see below). Adapted from (Binzegger et al., 2004), with long-range connections removed

**Presynaptic neurons**

| Postsynaptic neurons | | percent of cells | max no. synapses | P2/3 | B2/3 | nB2/3 | ss4(L4) | ss4(L2/3) | p4 | b4 | nb4 | p5(L2/3) | p5(L56) | b5 | nb5 | p6(L4) | p6(L56) | b6 |
|---|---|---|---|---|---|---|---|---|---|---|---|---|---|---|---|---|---|---|
| P2/3 | L2/3 | 26.3 | 5773 | 60.1 | 9.2 | 4.9 | 0.6 | 6.9 | 7.8 | 0.8 | - | 7.5 | - | - | - | 2.4 | - | - |
|  | L1 |  | 87 | 95.1 | 1.6 | - | - | 0.3 | 1.5 | 0.1 | - | 1.1 | - | - | - | - | - | - |
| B2/3 |  | 3.1 | 3702 | 53.7 | 11.0 | 12.3 | 0.5 | 6.1 | 6.8 | 0.9 | - | 6.6 | - | - | - | 2.1 | - | - |
| NB2/3 |  | 4.2 | 3144 | 57.8 | 12.6 | 4.6 | 0.5 | 6.6 | 7.5 | 0.9 | - | 7.2 | - | - | - | 2.2 | - | - |
| SS4(L4) |  | 9.3 | 4113 | 3.8 | 0.3 | - | 16.7 | 5.2 | 5.8 | 12.9 | 8.1 | 1.1 | 0.1 | - | - | 46.1 | - | - |
| SS4(L2/3) |  | 9.3 | 3610 | 7.7 | 0.6 | - | 15.6 | 5.3 | 5.9 | 12.8 | 7.6 | 1.5 | 0.1 | - | - | 43.0 | - | - |
| P4 | L4 | 9.3 | 3619 | 6.0 | 0.3 | - | 16.0 | 5.0 | 5.8 | 12.9 | 8.3 | 1.6 | 0.1 | 0.1 | - | 43.7 | 0.1 | - |
|  | L2/3 |  | 867 | 63.0 | 5.1 | 5.1 | 0.6 | 7.2 | 8.1 | 0.6 | - | 7.8 | - | - | - | 2.5 | - | - |
|  | L1 |  | 53 | 6.0 | 0.3 | - | 16.0 | 5.0 | 5.8 | 12.9 | 8.3 | 1.6 | 0.1 | 0.1 | - | 43.7 | 0.1 | - |
| B4 |  | 5.5 | 2359 | 8.0 | 0.7 | - | 15.1 | 5.2 | 5.8 | 14.8 | 7.3 | 1.5 | 0.1 | - | - | 41.6 | - | - |
| NB4 |  | 1.5 | 2636 | 3.7 | 0.3 | - | 16.3 | 5.1 | 5.6 | 14.8 | 7.9 | 1.1 | 0.1 | - | - | 45.0 | - | - |
| P5(L2/3) | L5 | 4.9 | 3971 | 49.9 | 2.0 | - | 3.6 | 2.2 | 8.2 | 1.0 | - | 12.7 | 1.1 | 2.1 | 12.5 | 2.5 | 2.2 | - |
|  | L4 |  | 198 | 4.0 | 0.1 | - | 17.4 | 5.4 | 6.0 | 9.5 | 8.4 | 1.1 | 0.1 | - | - | 48.0 | - | - |
|  | L2/3 |  | 413 | 62.9 | 5.1 | 5.1 | 0.6 | 7.2 | 8.1 | 0.6 | - | 7.8 | - | - | - | 2.4 | - | - |
|  | L1 |  | 12 | 97.7 | 1.7 | - | - | 0.3 | 1.5 | 0.1 | - | 1.1 | - | - | - | - | - | - |
| P5(L56) | L5 | 1.3 | 4588 | 49.3 | 1.8 | - | 3.6 | 2.2 | 8.1 | 0.9 | - | 12.5 | 1.3 | 2.0 | 13.0 | 2.5 | 2.8 | - |
|  | L4 |  | 666 | 4.0 | 0.1 | - | 17.4 | 5.4 | 6.0 | 9.5 | 8.4 | 1.1 | 0.1 | - | - | 48.0 | - | - |
|  | L2/3 |  | 1368 | 63.0 | 5.1 | 5.1 | 0.6 | 7.2 | 8.1 | 0.6 | - | 7.8 | - | - | - | 2.5 | - | - |
|  | L1 |  | 375 | 95.6 | 1.7 | - | - | 0.3 | 1.5 | 0.1 | - | 1.1 | - | - | - | - | - | - |
| B5 |  | 0.6 | 2744 | 49.5 | 2.5 | - | 3.6 | 2.2 | 8.1 | 1.2 | - | 12.6 | 1.1 | 2.3 | 12.3 | 2.5 | 2.2 | - |
| NB5 |  | 0.8 | 2744 | 49.5 | 2.5 | - | 3.6 | 2.2 | 8.1 | 1.2 | - | 12.6 | 1.1 | 2.3 | 12.3 | 2.5 | 2.2 | - |
| P6(L4) | L6 | 13.8 | 1326 | 6.1 | 0.3 | - | 1.8 | 2.1 | 3.1 | 0.2 | - | 0.2 | 12.0 | 0.9 | - | 2.9 | 32.4 | 38.1 |
|  | L5 |  | 979 | 51.0 | 0.9 | - | 3.7 | 2.3 | 8.4 | 0.6 | - | 13.0 | 1.1 | 1.6 | 12.7 | 2.5 | 2.3 | - |
|  | L4 |  | 1344 | 4.0 | 0.1 | - | 17.4 | 5.4 | 6.0 | 9.5 | 8.4 | 1.1 | 0.1 | - | - | 48.0 | - | - |
|  | L2/3 |  | 121 | 62.9 | 5.1 | 5.1 | 0.6 | 7.2 | 8.1 | 0.6 | - | 7.8 | - | - | - | 2.4 | - | - |
| P6(L56) | L6 | 4.6 | 2264 | 4.0 | 0.1 | - | 17.4 | 5.4 | 6.0 | 9.5 | 8.4 | 1.1 | 0.1 | - | - | 48.0 | - | - |
|  | L5 |  | 236 | 51.0 | 0.9 | - | 3.7 | 2.3 | 8.4 | 0.6 | - | 13.0 | 1.1 | 1.6 | 12.7 | 2.5 | 2.3 | - |
|  | L4 |  | 171 | 4.0 | 0.1 | - | 17.4 | 5.4 | 6.0 | 9.4 | 8.4 | 1.1 | 0.1 | - | - | 47.8 | - | - |
|  | L2/3 |  | 286 | 63.1 | 5.1 | 5.1 | 0.6 | 7.2 | 8.1 | 0.6 | - | 7.8 | - | - | - | 2.5 | - | - |
|  | L1 |  | 4 | 97.7 | 1.7 | - | - | 0.3 | 1.5 | 0.1 | - | 1.1 | - | - | - | - | - | - |
| B6 |  | 2.0 | 1310 | 6.1 | 0.3 | - | 1.8 | 2.1 | 3.1 | 0.2 | - | 0.2 | 12.0 | 0.9 | - | 2.9 | 32.4 | 38.1 |

**Table ESM4** Axonal arborisation radii for each neuron group in each layer (mm), adapted from Figure 8 of the supporting information of (Izhikevich and Edelman, 2008). Where no radius was given for a neuron group in a layer in which connections are specified in Table ESM3, we set the radius to 0.05 mm

|  | **L1** | **L2/3** | **L4** | **L5** | **L6** |
|---|---|---|---|---|---|
| **P2/3** | 0.55 | 1.12 | 0.15 | 1.00 | 0.15 |
| **B2/3** | 0.05 | 0.50 | 0.15 | 0.15 | 0.05 |
| **NB2/3** | 0.20 | 0.20 | 0.20 | 0.20 | 0.20 |
| **SS4(L4)** | 0.05 | 0.30 | 1.12 | 0.40 | 0.15 |
| **SS4(L2/3)** | 0.15 | 0.40 | 0.50 | 0.15 | 0.15 |
| **P4** | 0.15 | 1.12 | 0.15 | 0.55 | 0.15 |
| **B4** | 0.05 | 0.05 | 0.50 | 0.05 | 0.05 |
| **NB4** | 0.20 | 0.20 | 0.20 | 0.20 | 0.20 |
| **P5(L2/3)** | 0.15 | 0.40 | 0.30 | 0.50 | 0.25 |
| **P5(L5/6)** | 0.05 | 0.05 | 0.15 | 0.50 | 1.00 |
| **B5** | 0.05 | 0.05 | 0.05 | 0.5 | 0.05 |
| **NB5** | 0.20 | 0.20 | 0.20 | 0.20 | 0.20 |
| **P6(L4)** | 0.05 | 0.15 | 1.00 | 0.15 | 0.15 |
| **P6(L5/6)** | 0.05 | 0.05 | 0.15 | 0.50 | 1.00 |
| **B6** | 0.05 | 0.05 | 0.05 | 0.10 | 0.50 |

**Table ESM5** Synaptic weights (nS)

| | \multicolumn{15}{c}{**Presynaptic neurons**} |
|---|---|---|---|---|---|---|---|---|---|---|---|---|---|---|---|

| Postsynaptic neurons | P2/3 | B2/3 | NB2/3 | SS4(L4) | SS4(L2/3) | P4 | B4 | NB4 | P5(L2/3) | P5(L5/6) | B5 | NB5 | P6(L4) | P6(L5/6) | B6 |
|---|---|---|---|---|---|---|---|---|---|---|---|---|---|---|---|
| **P2/3** | 0.020 | 0.126 | 0.001 | 0.356 | 0.036 | 0.073 | 1.080 | - | 0.004 | - | - | - | 0.047 | - | - |
| **B2/3** | 0.560 | 0.026 | 0.001 | 0.701 | 0.078 | 0.161 | 0.228 | - | 0.074 | - | - | - | 0.159 | - | - |
| **NB2/3** | 0.408 | 0.069 | 0.014 | 0.872 | 0.085 | 0.173 | 0.581 | - | 0.159 | - | - | - | 0.178 | - | - |
| **SS4(L4)** | 0.001 | 0.043 | - | 0.067 | 0.092 | 0.061 | 0.010 | 0.003 | 0.069 | 0.069 | - | - | 0.004 | - | - |
| **SS4(L2/3)** | 0.001 | 0.043 | - | 0.067 | 0.092 | 0.061 | 0.011 | 0.003 | 0.069 | 0.069 | - | - | 0.004 | - | - |
| **P4** | 0.001 | 0.043 | 0.008 | 0.067 | 0.092 | 0.061 | 0.014 | 0.001 | 0.069 | 0.069 | - | - | 0.004 | 0.004 | - |
| **B4** | 0.101 | 0.098 | - | 0.627 | 0.627 | 0.627 | 0.025 | 0.003 | 0.841 | 0.841 | - | - | 0.062 | 0.062 | - |
| **NB4** | 0.139 | 0.244 | - | 0.318 | 0.318 | 0.318 | 0.068 | 0.013 | 1.058 | 1.058 | - | - | 0.052 | - | - |
| **P5(L2/3)** | 0.037 | 0.188 | 0.004 | 0.091 | 0.082 | 0.050 | 0.341 | 0.005 | 0.079 | 0.471 | 0.459 | 0.003 | 0.032 | 0.093 | - |
| **P5(L5/6)** | 0.037 | 0.188 | 0.004 | 0.091 | 0.082 | 0.050 | 0.289 | 0.005 | 0.062 | 0.335 | 0.416 | 0.003 | 0.032 | 0.093 | - |
| **B5** | 0.083 | 0.098 | - | 0.274 | 0.273 | 0.151 | 0.191 | - | 0.910 | 3.966 | 0.166 | 0.003 | 0.342 | 0.207 | - |
| **NB5** | 0.064 | 0.244 | - | 0.331 | 0.422 | 0.196 | 0.521 | - | 0.603 | 2.596 | 0.166 | 0.014 | 0.359 | 0.657 | - |
| **P6(L4)** | 0.003 | 1.045 | 0.015 | 0.137 | 0.145 | 0.095 | 0.226 | 0.001 | 0.084 | 0.055 | 0.293 | 0.004 | 0.064 | 0.062 | 0.075 |
| **P6(L5/6)** | 0.003 | 1.045 | 0.015 | 0.137 | 0.145 | 0.095 | 0.978 | 0.016 | 0.201 | 0.055 | 0.293 | 0.004 | 0.064 | 0.062 | 0.048 |
| **B6** | 0.123 | 0.140 | - | 0.274 | 0.273 | 0.151 | 0.193 | - | 0.091 | 0.091 | 0.021 | - | 1.105 | 0.768 | 0.015 |

**Table ESM6** Synaptic parameters (postsynaptic), from (Traub et al., 2005b)

|  | $E_{AMPA}$ (mV) | $E_{GABA}$ (mV) | $t_{AMPA}$ (ms) | $t_{GABA}$ (ms) |
|---|---|---|---|---|
| **P, SS** | 0 | -75 | 2.0 | 6.0 |
| **B, NB** | 0 | -75 | 0.8 | 3.0 |

**Table ESM7** Compartment IDs in each postsynaptic group that presynaptic neurons connect onto (compartment numbers illustrated in Fig. ESM2). Based on (Traub et al., 2005b)

|  | Presynaptic neurons | | | | | |
|---|---|---|---|---|---|---|
| Postsynaptic neurons | P2/3, P4 | SS | P5 | P6 | B | NB |
| **P2/3, P4** | 3,6,7,8 | 6-8 | 4,5 | 4 | 1,2,6 | 3-5,7,8 |
| **SS4** | 3,4,6,7 | 3,4,6,7 | 3,4,6,7 | 3,4,6,7 | 1,2,5 | 3,4,6,7 |
| **P5** | 2-9 | 2-5 | 2-5,7-9 | 2-5,7-9 | 1,2,7 | 3-5,7,8 |
| **P6** | 2-9 | 2,4,5 | 2-9 | 2-9 | 1,2,7 | 3-5,7,8 |
| **B** | 3,4,6,7 | 3,4,6,7 | 3,4,6,7 | 3,4,6,7 | 3,4,6,7 | 3,4,6,7 |
| **NB** | 3,4,6,7 | 3,4,6,7 | 3,4,6,7 | 3,4,6,7 | 3,4,6,7 | 3,4,6,7 |

**Table ESM8** Neocortical slice model layer boundaries. Brain surface is at 2600 microns, white matter boundary at 0 microns

| Layer | Upper boundary (μm) |
|---|---|
| 1 | 2600 |
| 2/3 | 2362 |
| 4 | 1835 |
| 5 | 1122 |
| 6 | 832 |

**Table ESM9** Random current input parameters

|  | Mean current (pA) | Standard deviation (pA) | Noise correlation time constant (ms) |
|---|---|---|---|
| **P2/3** | 360 | 110 | 2.0 |
| **B2/3** | 200 | 60 | 0.8 |
| **NB2/3** | 160 | 40 | 0.8 |
| **SS4** | 205 | 50 | 2.0 |
| **P4** | 250 | 70 | 2.0 |
| **B4** | 200 | 60 | 0.8 |
| **NB4** | 160 | 40 | 0.8 |
| **P5** | 860 | 260 | 2.0 |
| **B5** | 200 | 60 | 0.8 |
| **NB5** | 160 | 40 | 0.8 |
| **P6** | 660 | 170 | 2.0 |
| **B6** | 200 | 60 | 0.8 |

**Table ESM10** List of the extra cell models downloaded from NeuroMorpho.org (Ascoli et al., 2007) used to investigate LFP range and magnitude in Fig. 1 and Fig. 2. Further experimental details are given in the Supplementary Methods, below

| Type | NeuroMorpho.Org ID | Species/Region | Original publication |
|---|---|---|---|
| P2/3 | NMO_00850 | Cat visual cortex | Kisvárday and Eysel, 1992 |
| P2/3 | NMO_00851 | Cat visual cortex | Kisvárday and Eysel, 1992 |
| P2/3 | NMO_00853 | Cat visual cortex | Kisvárday and Eysel, 1992 |
| P2/3 | NMO_00854 | Cat visual cortex | Kisvárday and Eysel, 1992 |
| P2/3 | NMO_00859 | Cat visual cortex | Kisvárday and Eysel, 1992 |
| P2/3 | NMO_00856 | Cat visual cortex | Kisvárday and Eysel, 1992 |
| P2/3 | NMO_00857 | Cat visual cortex | Kisvárday and Eysel, 1992 |
| P2/3 | NMO_00935 | Cat visual cortex | Hirsch et al., 2002 |
| P2/3 | NMO_05813 | Cat suprasylvian gyrus | Volgushev et al., 2006 |
| P2/3 | NMO_05811 | Cat suprasylvian gyrus | Volgushev et al., 2006 |
| P5 | NMO_00880 | Cat temporal sulcus | Contreras et al., 1997 |

**Table ESM11** Firing rates in the model under the four conditions illustrated in Fig. 8: model as described, exhibiting gamma oscillation; model with P2/3 -> B2/3 connections at 1% baseline strength; model with B2/3 -> P2/3 connections at 1% baseline strength; model with increased input current to B2/3 neurons (150% mean and standard deviation)

|  | Baseline | 1% P2/3 -> B2/3 weight | 1% B2/3 -> P2/3 weight | 150% input to B2/3 |
|---|---|---|---|---|
| P2/3 | 3.6 | 7.0 | 9.4 | 0.0 |
| B2/3 | 24.8 | 1.1 | 60.5 | 66.8 |
| NB2/3 | 4.2 | 8.4 | 4.0 | 0.0 |
| SS4(L4) | 1.3 | 1.4 | 1.0 | 1.0 |
| SS4(L2/3) | 1.2 | 1.4 | 0.7 | 0.6 |
| P4 | 1.6 | 1.9 | 1.0 | 0.9 |
| B4 | 5.6 | 9.2 | 3.4 | 1.2 |
| NB4 | 2.1 | 2.6 | 1.4 | 0.6 |
| P5(L2/3) | 3.9 | 4.3 | 2.6 | 2.0 |
| P5(L5/6) | 3.6 | 4.3 | 2.1 | 1.4 |
| B5 | 16.8 | 23.4 | 4.0 | 0.3 |
| NB5 | 3.3 | 4.8 | 0.1 | 0.0 |
| P6(L4) | 1.5 | 2.3 | 0.7 | 0.5 |
| P6(L5/6) | 1.4 | 1.9 | 0.7 | 0.6 |
| B6 | 7.1 | 9.3 | 5.7 | 4.5 |

## SUPPLEMENTARY METHODS

**Reduced neuron models**
During our investigations with the reduced cell models, we found some inconsistencies when comparing their dynamics with the original cell reconstructions when using the reduced compartment dimensions given in (Bush and Sejnowski, 1993). We therefore recalculated the compartment lengths and diameters from the three cell types specified in (Mainen and Sejnowski, 1996) using the method specified in (Bush and Sejnowski, 1993). For these calculations, we used a version of the NEURON code originally written by Alain Destexhe to reduce a compartmental model to 3 compartments (Destexhe et al., 1998), modified by Michael Hines to work for any number of compartments. This code implements the method described in (Bush and Sejnowski, 1993). Michael Hines' version of this code is available on the NEURON forum at http://www.neuron.yale.edu/phpbb/viewtopic.php?f=13&t=589. The recalculated cell dimensions are given in Table ESM1. We used our recalculated dimensions both for the LFP simulation comparison between the cell reconstructions and reduced cell models and for the necortical slice model.

**Comparison of LFPs from reduced & full neuron morphologies**
Fig. 1 and Fig. 2 show comparisons between LFPs calculated using populations containing reduced and full neuron models, and include further comparisons with extra (full-scale) neurons to place the results in the context of general biological variability (experiments described in the main text). These extra neurons were downloaded from NeuroMorpho.Org (Ascoli et al., 2007) and are listed in Table ESM10, above. Prior to being simulated in LFPy, we first removed all axonal compartments and rotated the neurons so that their apical dendrites were approximately parallel with the $z$-axis.

**Connectivity**
The connectivity model for the neocortical slice model (implemented in VERTEX) is described in the main text; here we give details of how neurons spanning multiple layers are connected. Pyramidal neuron dendrites span several layers above their soma layer, and connectivity statistics are provided per layer for pyramidal cells in (Binzegger et al., 2004) – see Table ESM3. As all neurons within a population are the same size, but have different soma positions, each neuron's compartments will cross the model's layer boundaries at different points. For simplicity, we ignore this variability for the purposes of connecting up the model, defining the layers in which a compartment resides based on its position when its neuron's soma is positioned in the centre of its layer. If several compartments could be chosen, then the compartment on which the synapse is made is chosen randomly, with each possible compartment having a probability of being selected equal to the membrane area of the compartment in the layer divided by the neuron's total membrane area in the layer. The chosen compartment must also be allowed according to Table ESM7. For our simulations, we allowed multiple synapses between a single pre- and postsynaptic neuron pair (targets randomly chosen with replacement), but did not allow autapses. These options can be configured in the simulation parameters.

**Neuron and synapse dynamics**
For the dynamics simulation of the neocortical slice model, we chose the 2 variable adaptive exponential (AdEx) model (Brette and Gerstner, 2005). The AdEx model can reproduce most of the dynamical features exhibited by cortical neurons (Naud et al., 2008), all its parameters have a direct biological correlate (Gerstner and Brette, 2009) making the model easy to interpret and modify in light of new experimental data, its sub-threshold behaviour is realistic (Badel et al., 2008), and its bifurcation structure is well characterised and is the same as the commonly used Izhikevich model (Naud et al., 2008; Touboul and Brette, 2008). It can be extended to include passive dendrite compartments (Clopath et al., 2007; Gerstner and Brette, 2009) required for LFP simulation, which we did by incorporating the AdEx dynamics into the somatic compartment of the passive cell model reductions described above. A modification of the AdEx model was also used recently in another study of gamma oscillations (Economo and White, 2012).

Each neuron is modelled as a passive cable structure of cylindrical compartments, with the AdEx spiking mechanism at the soma compartment. Passive parameters are given above. The somatic membrane potential $v_s$ evolves according to equation ESM1:

$$C_s \frac{dv_s}{dt} = -g_{leak,s}(v_s - E_{leak}) - \sum_j g_{sj}(v_s - v_j) + g_{leak,s}\Delta_t \exp\left(\frac{v_s - V_t}{\Delta_t}\right) - w + I_s, \quad \text{(ESM1)}$$

$$\tau_w \frac{dw}{dt} = \alpha(v_s - E_{leak}) - w,$$

if $v_s \geq v_{cutoff}$:

$\quad\quad v_s \leftarrow v_{reset}$,

$\quad\quad w \leftarrow w + \beta$,

where $C_s$ is the soma membrane capacitance, $g_{leak,s}$ is the soma leak conductance (= reciprocal of soma membrane resistance), $E_{leak}$ is the leak reversal potential, $g_{sj}$ are the conductances between the soma and its $j$ connected compartments, $v_j$ is the membrane potential of the $j^{th}$ connected compartment, $\Delta_t$ is a constant defining the spike steepness, $V_t$ is the instantaneous threshold potential, $w$ is a current representing the combined slow ionic currents, $I_s$ is the total current input at the soma (from synaptic and externally applied currents), $\tau_w$ is the time constant of the slow current $w$, $\alpha$ is the scale factor of the slow current, $v_{cutoff}$ is the potential at which a spike is said to have been fired, $v_{reset}$ is the membrane potential to which $v_s$ returns after a spike, and $\beta$ is the instantaneous change in the value of the slow current $w$ after a spike (Bretteand Gerstner, 2005). All dendrites are passive:

$$C_k \frac{dv_k}{dt} = -g_{leak,k}(v_k - E_{leak}) - \sum_j g_{kj}(v_k - v_j) + I_k, \quad \text{(ESM2)}$$

where the symbols are as before, for dendritic compartment $k$ rather than soma $s$.

We adjusted the parameters for each neuron type to produce similar spiking patterns to the model neurons described in (Traub et al., 2005b). Each cell type's passive parameters were defined by its morphology and the electrotonic parameters given in Table ESM2; therefore, the parameters adjusted to fit the spiking responses of the Traub neurons were the spike slope factor $\Delta_t$, threshold $V_t$, adaptation time constant $\tau_w$, adaptation coupling parameter $\alpha$, reset value $v_{reset}$, and instantaneous adaptation current increase $\beta$. We employed a qualitative approach to parameter adjustment, guided by the analysis of the AdEx model in (Naud 2008). According to the classifications in (Naud 2008), B, SS and P6 cells have a sharp reset, while NB, P2/3, P4 and P5 cells have a broad reset. B cells are non-adapting; SS and P6 cells are adapting; P2/3, P4, P5 and NB cells show an initial burst. These properties were chosen based on the membrane potential traces reported in Appendix A of (Traub et al., 2005b).

The model includes AMPA and GABA$_A$ synapses, each modelled as single exponential, conductance-based synapses. When a neuron fires a spike, the synaptic conductances ($g_{AMPA}$ for excitatory presynaptic neurons, $g_{GABA}$ for inhibitory presynaptic neurons) at the contacted target compartments $k$ (specified in the connectivity matrix) are increased by the synaptic weight (specified in the weights matrix) after the relevant axonal delay time, then decay exponentially:

$$\frac{dg_{AMPA,k}}{dt} = -\frac{g_{AMPA,k}}{\tau_{AMPA}}, \quad \text{(ESM3)}$$

$$\frac{dg_{GABA,k}}{dt} = -\frac{g_{GABA,k}}{\tau_{GABA}},$$

where $\tau_{AMPA}$ and $\tau_{GABA}$ are the AMPA and GABA$_A$ synaptic decay constants, respectively (specified in Table ESM6). As we assume that the conductances of individual synapses sum linearly in each compartment, we only need one variable per type of synaptic conductance per compartment, rather than keeping track of all synapses individually. The total synaptic current $I_k$ at compartment $k$ at time $t$ is then given by

$$I_k(t) = g_{AMPA,k}(v_k(t) - E_{AMPA}) + g_{GABA,k}(v_k(t) - E_{GABA}), \quad \text{(ESM4)}$$

where $E_{AMPA}$ and $E_{GABA}$ are the reversal potentials for AMPA and GABA$_A$, respectively. Synaptic weights (Table ESM5) were chosen based on those reported in (Traub et al., 2005b), scaling the weights according to the number of synapses between groups in our model compared with the Traub model. Our neuron populations did not match theirs exactly, with the following differences (in addition to different numbers of neurons and synapses): our model includes interneurons in every layer, while the Traub model has only "superficial" and "deep" interneurons (with the deep interneurons providing inhibition to layer 4); the Traub model only has spiny stellate cells in layer 4 (no pyramidal or interneurons); the Traub model contains fast rhythmic bursting pyramidal cells in layer 2/3 and intrinsically bursting pyramidal cells in layer 5 – our model contains no bursting neurons; our model contains synapses between some neuron groups that are not present in the Traub model. We therefore had to make several arbitrary decisions when setting some synapse weights between groups.

**Model input**
We stimulate our model to mimic the bath application of kainate. This stimulates the pyramidal cell axonal plexus, providing the neurons with excitatory drive. We simulate this by applying independent random input currents $I_{Ki}$ to each neuron $i$, modelled as Ornstein-Uhlenbeck processes similar to (Arsiero et al., 2007) using Gillespie's exact discretisation method:

$$I_{Ki}(t+\delta t) = I_{Ki}(t) + \left(1 - \exp\left(\frac{-\delta t}{\tau_K}\right)\right) \times \left(m_{Ki} - I_{Ki}(t)\right) + \sqrt{1 - \exp\left(-2\frac{\delta t}{\tau_K}\right)} \times S_{Ki} \times N_{Ki}(0,1),$$  (ESM5)

(Gillespie, 1996), where $\delta t$ is the length of the time step, $\tau_k$ is the noise correlation time constant, $m_{Ki}$ is the mean current value, $S_{Ki}$ is the standard deviation and $N_{Ki}(0,1)$ is a normally distributed random number. The random current is distributed across the neuron's compartments proportionally to the compartment membrane areas. Any currents falling below zero are reset to zero for that time step, so that the input current is always either positive or zero.

LFPs are calculated by summing the membrane currents of each compartment, weighted by distance from the electrode tips (described in the main text). The membrane current $I_{mem,k}$ of compartment $k$ is just the negative of the axial current $I_{ax,k}$ entering the compartment (Johnston and Wu, 1995):

$$I_{mem,k} = \sum_j g_{kj}(v_k - v_j) = -I_{ax,k}.$$  (ESM6)

**Model implementation**
The neocortical slice model was implemented in our VERTEX simulation tool. VERTEX is written in Matlab, using the Matlab Parallel Computing Toolbox for parallelisation, though it can also be run serially. It is designed to be easy to use: neuron groups, connectivity patterns, model size/layers and simulation settings are defined in Matlab structures, requiring minimal programming ability to specify models of different cortical areas, explore the parameter space, or apply different stimuli.

First, the model is initialised by distributing neurons across parallel processes, positioning the neurons, setting up the connectivity matrix, and calculating axonal delays. Next, the electrode locations are specified and distances between each electrode and each compartment are calculated, using either the point distance for somas, or the line source distance for dendrites. Pre-calculating the constant values used in the field potential calculations minimises the impact of calculating the LFP during the simulation.

We used the methods outlined in (Morrison et al., 2005) for parallel simulation. These minimise communication overhead by storing synapse information (delays, postsynaptic neuron IDs and compartment IDs) on the postsynaptic side, so only spiking presynaptic IDs and timestamps need to be exchanged between processes. Spikes do not need to be delivered every time step: assuming the minimum synaptic delay is $d_{min} \cdot h$ (where $h$ is the step size), spikes can be buffered on the pre-synaptic side and exchanged every $d_{min}$ time steps. Processes communicate using the complete pairwise exchange algorithm (Tam & Wang, 2000; Morrison et al., 2005).

The simulation made use of vectorised data structures and algorithms in order to keep run-times reasonable (Brette and Goodman, 2011). The benefits of vectorisation increase with model size as fewer interpretation overheads are incurred per variable. In addition to the methods described in (Brette and Goodman, 2011), which do not include compartmental neuron models, we vectorise the calculation of the axial currents between compartments.

**SUPPLEMENTARY REFERENCES**